\documentclass[preprint,aps,amsmath, nofootinbib,floatfix,preprintnumbers]{revtex4}
 
\usepackage{graphicx}
\usepackage{cancel}
\usepackage{amssymb}
\usepackage{textcomp}
\usepackage{amsmath}
\usepackage{bm}
\usepackage{times}
\usepackage{epsfig}
\usepackage{color}
\usepackage{subfigure}

\usepackage{comment}

\def\met{{\cancel{E}_{T}}}
 \newcommand{\gev}{{\rm GeV}}

\def\s1{\ensuremath{\tilde t_1}}
\def\b1{\ensuremath{\tilde b_1}}

\newcommand{\gsim}{\buildrel > \over {_\sim}}

\newcommand{\minigraph}[5][0.25in]{\begin{minipage}{#2}\begin{center}\includegraphics[width=#2]{#5}\\\vspace{#3}\hspace{#1}{\footnotesize #4}\end{center}\end{minipage}}

\begin{document}

\preprint{ACFI-T14-08}



\title{Impact of LSP Character on Slepton Reach at the LHC}
\author{
         Jonathan Eckel$^{1}$\footnote{eckel@physics.arizona.edu},
         Michael J. Ramsey-Musolf$^{2,3}$\footnote{mjrm@physics.umass.edu},
         William Shepherd$^{4,5}$\footnote{wshepher@ucsc.edu},
         Shufang Su$^{1}$\footnote{shufang@email.arizona.edu} }

\affiliation{
$^{1}$ Department of Physics, University of Arizona, Tucson, AZ  85721\\
$^{2}$ Amherst Center for Fundamental Interactions, Department of Physics University of Massachusetts-Amherst, Amherst, MA 01003\\
$^{3}$ Kellogg Radiation Laboratory, California Institute of Technology, Pasadena, CA 91125\\
$^{4}$ Department of Physics, University of California - Santa Cruz, Santa Cruz, CA 95064\\
$^5$ Santa Cruz Institute for Particle Physics, Santa Cruz, CA 95064
}

\begin{abstract}
 
Searches for supersymmetry at the Large Hadron Collider (LHC) have significantly constrained the parameter space associated with colored superpartners, whereas the constraints on color-singlet superpartners are considerably less severe.   In this study, we investigate the dependence of slepton decay branching fractions on the nature of the lightest supersymmetric particle (LSP).  In particular, in the Higgsino-like LSP scenarios, both decay branching fractions of $\tilde\ell_L$ and $\tilde\nu_\ell$ depend strongly on the sign and value of $M_1/M_2$, which has strong implications for the reach of dilepton plus $\met$ searches for slepton pair production. We extend the experimental results for same flavor, opposite sign dilepton plus $\met$ searches at  the 8 TeV LHC to various LSP scenarios.  We find that the LHC bounds on sleptons are strongly enhanced for a non-Bino-like LSP: the 95\% C.L. limit for $m_{\tilde\ell_L}$ extends from 300 GeV for a Bino-like LSP to about 370 GeV for  a Wino-like LSP. The bound for $\tilde\ell_L$ with a Higgsino-like LSP is the strongest ($\sim$ 490 GeV) for $M_1/M_2\sim -\tan^2\theta_W$ and is the weakest ($\sim$  220 GeV) for $M_1/M_2\sim \tan^2\theta_W$.    We also calculate prospective slepton search reaches at the 14 TeV LHC. With 100 $\rm fb^{-1}$ integrated luminosity,  the projected 95\% C.L. mass reach for the left-handed slepton varies from 550 (670) GeV for a Bino-like (Wino-like) LSP to 900 (390) GeV for a Higgsino-like LSP under the most optimistic (pessimistic) scenario. The reach for the right-handed slepton is about 440 GeV. The corresponding 5$\sigma$ discovery sensitivity is about 100 GeV smaller. For 300 $\rm fb^{-1}$ integrated luminosity, the reach is about 50 $-$ 100 GeV higher.  
 \end{abstract}

\maketitle

\section{Introduction}
\label{sec:intro}

While the discovery of a Standard Model (SM)-like Higgs  boson at 125 GeV has been the most significant result obtained at the Large Hadron Collider (LHC) to date \cite{aad:2012gk,Chatrchyan:2012ufa},  no signal for new physics beyond the SM has yet emerged. Any new colored particle would be the best targets for the LHC due to the large QCD production cross sections. Searches for hadronic final states do, however, suffer from the complicated hadronic environment. Hadronically-quiet new physics searches in leptonic final states are typically challenging due to the smaller electroweak production cross sections, yet the associated SM backgrounds are more clearly understood.

Weak scale supersymmetry (SUSY) is one of the most promising new physics scenarios, and the search for supersymmetric particles continues to be one of the main efforts of LHC studies. LHC SUSY searches have largely focused on  gluinos and squarks.  The null results have set lower limits of about 1200 GeV and 800 GeV, respectively, for the masses of gluinos and  degenerate first- and second- generation squarks \cite{SUSYcolor}. The limits on the electroweak sector of the Minimal Supersymmetric Standard Model (MSSM), however, are much less stringent.

If low energy supersymmetry is realized in the nature, sleptons are
likely to be light.  This feature emerges in  the Gauge Mediated SUSY-breaking 
scenarios \cite{GMSB} and the Anomaly Mediated SUSY-breaking
scenarios \cite{AMSB}, wherein the slepton masses are proportional to
the electroweak gauge couplings.  Even in the  minimal Gravity Mediated SUSY-breaking scenarios (mSUGRA)
\cite{mSUGRA} where all the scalars have a common mass $m_0$ at a high
energy input scale, renormalization group running to low energies typically
pushes up the squark mass (due to the contributions of strongly
interacting gluinos) while the sleptons remain relatively light. The observation of
sleptons, even in the presence of the strong lower bounds on squark and gluino masses, would be 
consistent with these expectations. Thus, it is timely to fully explore
the discovery potential of the LHC for the lepton superpartners.

In the $R$-parity conserving MSSM,  the lightest neutralino   $\chi_1^0$ 
can be a natural candidate for
Weakly Interacting Massive Particle (WIMP) dark matter \cite{neutralinoDM} when it is the LSP.  When sleptons are light, the
$t$-channel process $\chi_1^0 \chi_1^0\to \ell^+\ell^-$   mediated by the exchange of sleptons can be important
in determining the $\chi_1^0$ annihilation cross section \cite{sleptonrelic}, and for fairly degenerate spectra of sleptons and $\chi_1^0$, coannihilation processes can also become important \cite{coann}.  Therefore, discovery of the sleptons would
not only provide a verification of low energy supersymmetry in nature; precise
measurement of their masses could also play an important role in
determining the relic density of  the neutralino LSP.  

Light sleptons also contribute to  low energy precision observables, such as the electron and proton weak charges that can be measured in parity-violation $ee$ M\o ller scattering and $ep$ scattering \cite{moller, qweak}, respectively, the muon anomalous magnetic moment \cite{muong-2}, or tests of first row CKM unitarity \cite{CKM}.  With the precision achieved (attainable) in current (future) measurements\cite{Kumar:2013qya,Albrecht:2013wet}, these low energy observables  provide an indirect probe of the slepton sector that complements the LHC direct search.

Earlier studies of the slepton discovery potential at the LHC 
focused primarily on the Drell-Yan pair production of slepton pairs, with
each slepton decaying  directly to a lepton and $\chi_1^0$ \cite{delAguila:1990yw, Baer:1993ew,
  Andreev:2004qq}.  Most of those studies have been performed either in the mSUGRA
framework or for a certain set of benchmark points only.  Dilepton plus missing $E_T$ final states have also been searched for at the LHC. When the results are interpreted in terms of slepton Drell-Yan pair production with direct decays to a Bino-like LSP, the current limit is fairly weak: $m_{\tilde\ell_L}\gsim$ 300 GeV for left-handed sleptons (${\tilde\ell}_L$) with a relatively light LSP \cite{sleptonATLAS,sleptonCMS}.
 
Sleptons can also be produced in the cascade decay of gauginos  when  kinematically accessible.  
The gaugino pair production cross sections are typically larger than that of the direct slepton Drell-Yan process, given the fermionic nature of the gauginos.  Once sleptons are lighter than gauginos, the gaugino dominantly decays to a slepton and lepton, with the slepton subsequently decaying to another lepton and the LSP.  For heavier neutralinos and charginos, such lepton-rich final states  greatly extends the reach of neutralino and charginos at the LHC~\cite{Aad:2014vma,Khachatryan:2014qwa}. In addition, imposing a sharp cut on the invariant mass distribution of two leptons produced in the $\chi_2^0$ decay could provide   further discrimination of the signal from the SM backgrounds, potentially allowing for discovery of the slepton in gaugino decays \cite{Eckel:2011pw}.

The implications of null results in the searches of neutralino/chargino decay via sleptons, however, are limited.  First, such experimental searches apply only to the case when sleptons are lighter than heavier gauginos; naturally there is no sensitivity to sleptons from gaugino decays once the decay is kinematically forbidden.  Second, even when sleptons are lighter than heavier gauginos, the experimental limits apply only to the case of Wino-like pair-produced gauginos with a Bino-like LSP and are, therefore, only sensitive to the stau or the left-handed slepton (${\tilde\ell}_L$). Finally, when the $\chi_2^0$ and $\chi_1^\pm$ are Higgsino-like states, no reach in the slepton mass can be derived even if the cascade decay is kinematically allowed, since the branching fraction into sleptons is highly suppressed by the small lepton Yukawa couplings 
and the small gaugino fractions of the the neutralino and chargino states.
 
Due to these limitations for the production of sleptons via neutralino/chargino decays, we are motivated to investigate the reach for sleptons via direct slepton Drell-Yan pair production, focusing on same flavor, opposite sign dilepton plus   $\met$ signal.   Earlier studies of the slepton searches at the LHC \cite{delAguila:1990yw,Baer:1993ew,Andreev:2004qq}  assumed a Bino-like LSP.  The sensitivity of this channel, however, depends sensitively on the   slepton being either left- or right-handed, as well on the composition of the LSP as being either Bino, Wino, or Higgsino dominated.  Utilizing the current search channel of dilepton plus $\met$ with  data collected at the   8 TeV LHC, we re-interpret the results that have been   presented by the ATLAS and CMS collaborations assuming a Bino-like LSP for cases with a Wino-like or a Higgsino-like LSP. We also study the   exclusion limits and discovery reach for sleptons at the 14 TeV LHC for various choices of the LSP.

The outline of the paper is as follows.  In Sec.~\ref{sec:MSSMslepton}, we
give a brief review of the slepton sector in the MSSM and discuss its dominant production and decay channels for various slepton and neutralino/chargino spectra.
In Sec.~\ref{sec:limit}, we summarize the current limits on the slepton searches,  from both LEP searches and the latest LHC results. In Sec.~\ref{sec:7TeV}, we interpret the ATLAS results  on the opposite sign dilepton plus $\met$ search (which assume a Bino-like LSP) in the cases of Wino-like and Higgsino-like  LSP, including additional production from sneutrinos in the case of the $\tilde\ell_L$ as well.  
In Sec.~\ref{sec:collider},  we study the reach for sleptons at the 14 TeV LHC.  In Sec.~\ref{sec:conclusion},   we conclude. 

\section{Sleptons in the MSSM}
\label{sec:MSSMslepton}

\subsection{Slepton spectrum}
The LHC slepton sensitivity considered here depends on both the slepton  pair production cross sections and the detailed nature of the branching fractions for the slepton decays. The latter, in turn, is determined by the electroweakino (chargino/neutralino)  spectrum. For simplicity, we consider the low-lying spectrum of the MSSM electroweak sector to include only sleptons, neutralinos and charginos.  We also assume negligible flavor mixing between the slepton generations and zero left-right mixing of the first two generation sleptons (motivated by their small Yukawa couplings). We can then label the charged slepton mass eigenstates for the first two generations as $\tilde\ell_L$ and $\tilde\ell_R$, for $\ell=e,\ \mu$, with masses $m_{\tilde\ell_L}$ and $m_{\tilde\ell_R}$, respectively.  
These masses are governed by the soft breaking mass terms $m_{SL}$ and $m_{SR}$:
$m_{\tilde\ell_L}^2 = m_{SL}^2 + \Delta_{\tilde\ell_L}$ and $m_{\tilde\ell_R}^2 = m_{SR}^2 + \Delta_{\tilde\ell_R}$, where the D-term contributions are $\Delta_{\tilde\ell_L} = (-\frac{1}{2} - \sin^2{\theta_W}) m_Z^2\cos 2 \beta$ and $\Delta_{\tilde\ell_R} = - \sin^2{\theta_W}m_Z^2\cos 2 \beta$.    The sneutrino masses are controlled by $m_{SL}$ as well and are, therefore, related to $m_{\tilde\ell_L}$ with a small splitting introduced by electroweak effects: $m_{\tilde\nu_\ell}^2 = m_{\tilde\ell_L}^2 + m_W^2\cos 2 \beta $;  for the  range of $\tan\beta > 1$, $m_{\tilde\nu_\ell} < m_{\tilde\ell_L}$. The phenomenology and implication of sizable flavor mixing in the slepton sector can be found in Ref.~\cite{sleptonflavor,Krasnikov:1996np}.   For the third generation charged leptons (staus),   left-right mixing may be sizable, especially if $\tan\beta$ is large. We focus here on the  first two generations of sleptons, although our approach could be  adapted to the stau case as well by taking the tau tagging efficiency and stau left-right mixing into account.   

The decay of sleptons depends on the composition and spectrum of neutralinos and charginos, which is set mainly by the Bino, Wino, and Higgsino mass parameters  $M_1$, $M_2$ and $\mu$, respectively.   We consider three representative cases:
\begin{itemize}
\item Bino-like LSP:  $|M_1|<|M_2|, \ |\mu|$, yielding a neutralino LSP $\chi_1^0$ that is Bino-like.    
\item Wino-like LSP:  $|M_2|<|M_1|, \ |\mu|$, yielding a Wino-like LSP  $\chi_1^0$  degenerate with $\chi_1^\pm$. 
\item Higgsino-like LSP:   $|\mu|<|M_1|,\ |M_2|$, yielding a Higgsino-like LSP $\chi_1^0$ degenerate with $\chi_2^0$ and $\chi_1^\pm$.  
\end{itemize}
In the Wino-like LSP and Higgsino-like LSP cases, $\chi_1^\pm$ (and $\chi_2^0$ in the Higgsino-like LSP case) decays to the neutralino LSP with very soft jets and leptons. Identifying these decays is very difficult at the LHC. For these cases, then, the nearly degenerate neutralino and chargino states all appear as $\met$ at the LHC.

In our discussion below, we assume the slepton  decays directly  to the $\chi_1^0$ LSP (and neutralino/chargino states that are degenerate with the LSP for the Wino- or Higgsino-like LSP cases) plus one lepton, a mode that is most likely to occur when the slepton is lighter than all other heavier neutralinos and charginos.   In cases when sleptons are heavier than charginos and neutralinos other than the LSP (and its nearly degenerate neutralino/chargino states),  sleptons may decay into those neutralino/chargino states, which subsequently cascade decay to the LSP.  The final states from such processes are typically more complicated, involving multi-leptons, multi-jets and $\met$.  While a slepton search relying on such slepton cascade decays is complementary to the one assuming direct decay of the slepton to the LSP plus a lepton, an analysis of the cascade decay scenario goes beyond the scope of our current study, and we leave it for future work.  

\subsection{Slepton decays}

We now turn to the slepton branching fractions  for the three different LSP cases. For the Bino-like LSP,  $\tilde{\ell}_L$ and $\tilde\ell_R$ both decay to $\ell \chi_1^0$, and $\tilde{\nu}$ decays to $\nu \chi_1^0$ with 100\% branching fraction.   For the Wino-like LSP, $\tilde{\ell}_L$ decays to $\ell \chi_1^0, \nu \chi_1^\pm$ ($\tilde{\nu}_L$ decays to  $\nu \chi_1^0, \ell \chi_1^\pm$) with branching fractions of 33\% and 67\%, respectively. These branching fractions are set by the $\sqrt{2}$ enhancement of charged current coupling relative to that of the neutral current. The $\tilde\ell_R$ decays to $\ell \chi_1^0$ with a branching fraction of nearly 100\% via a small Wino$-$Bino mixing. The decay of $\tilde{\ell}_R$ to $\nu \chi_1^\pm$ is highly suppressed by the small lepton Yukawa couplings.  

For the Higgsino-like LSP case, due to the strong suppression of the small lepton Yukawa coupling, $\tilde{\ell}_L$ and  $\tilde{\nu}_L$ decay to $\chi_{1,2}^0$ and $\chi_1^\pm$ via the Bino- and Wino-components of $\chi_{1,2}^0$ and $\chi_1^\pm$.  The branching fractions to $\chi_{1,2}^0$ depend on the relative   Bino and Wino fractions of the   $\chi_{1,2}^0$:  $|N_{i\tilde{B}}|^2$ and $|N_{i\tilde{W}}|^2$ ($i=1,2$), respectively, which are given to leading order in $m_Z/(M_{1,2}\pm \mu)$ by: 
\begin{eqnarray}
 N_{1\tilde{B}}=(s_\beta + c_\beta)\frac{s_W m_Z}{\sqrt{2}(M_1-\mu)}
  \qquad N_{2\tilde{B}}= -(s_\beta - c_\beta)\frac{s_W m_Z}{\sqrt{2}(M_1+\mu)}\\
N_{1\tilde{W}}=-(s_\beta + c_\beta)\frac{c_W m_Z}{\sqrt{2}(M_2-\mu)} \qquad N_{2\tilde{W}}=(s_\beta - c_\beta)\frac{c_W m_Z}{\sqrt{2}(M_2+\mu)}
\end{eqnarray}
where $s_W=\sin\theta_W$, $c_W=\cos\theta_W$ for $\theta_W$ being the weak mixing angle;  $s_\beta=\sin\beta$ and $c_\beta=\cos\beta$.  In   arriving at these expressions, we have assumed that $|M_{1,2}-\mu|\gg m_Z$. Note that the relative sign between the Bino and Wino components of the neutralinos is physical, and has interesting consequences.    Similarly, the Wino fractions of $\chi_1^\pm$ are given by the absolute squares of
\begin{equation}
U_{1\tilde{W}^-}=(c_\beta + s_\beta \frac{\mu}{M_2}) \frac{\sqrt{2} c_W m_Z}{M_2}, \qquad
V_{1\tilde{W}^+}=(s_\beta + c_\beta \frac{\mu}{M_2}) \frac{\sqrt{2} c_W m_Z}{M_2}.
\end{equation}
Note that we have explicitly kept the sub-leading term $\mu/M_2$ in the mixing coefficient since it can be important for the case of large $\tan\beta$ (as $c_\beta$ goes to zero) in $U_{1\tilde{W}^-}$,  which is relevant for $\tilde\ell_L$ decays. 

The partial decay widths for the charged slepton and sneutrino decays into Higgsino-like LSPs are given approximately by  
\begin{eqnarray}
\Gamma(\tilde\ell \to\ell \chi_{1,2}^0)&=& C \ (s_\beta\pm c_\beta)^2 
( m_Z\frac{  s_W^2}{M_1 \mp \mu}-m_Z\frac{c_W^2}{M_2 \mp \mu})^2,  \label{eq:slL_chi10}\\
\Gamma(\tilde\ell \to\nu_\ell  \chi_{1}^\pm)&=& C \ 8 c_W^4  (c_\beta+ s_\beta \frac{\mu}{M_2})^2 
( \frac{m_Z}{M_2} )^2,  
\label{eq:slL_chi1pm}\\
\Gamma(\tilde\nu_\ell \to\nu_\ell \chi_{1,2}^0)&=& C \ (s_\beta\pm c_\beta)^2 
( m_Z\frac{  s_W^2}{M_1 \mp \mu}+m_Z\frac{c_W^2}{M_2 \mp \mu})^2,  \label{eq:snuL_chi10}\\
\Gamma(\tilde\nu_\ell \to \ell  \chi_{1}^\pm)&=& C \ 8 c_W^4  (s_\beta+ c_\beta \frac{\mu}{M_2})^2 
( \frac{m_Z}{M_2} )^2 , 
\end{eqnarray}
where  
\begin{equation}
C=\frac{1}{16 \pi}\frac{e^2}{4 s_W^2 c_W^2}   \frac{(m_{\cal P}^2 - m_{\cal D}^2)^2}{m_{\cal P}^3}
\end{equation}
for $m_{\cal P}$ and $m_{\cal D}$ being the  the parent slepton mass and daughter neutralino/chargino mass, respectively. 
The ``$\pm$" in Eqs.~(\ref{eq:slL_chi10}) and (\ref{eq:snuL_chi10})  correspond to $\chi_{1}^0$ and $\chi_2^0$, respectively.  Given the near degeneracy of $\chi_{1,2}^0$ for the Higgsino states, the rates for decays to these two channels are usually added together since $\chi_{1,2}^0$ both appear as $\met$ at hadron colliders.  In the limit of $|\mu| \ll |M_{1,2}|$, 
\begin{eqnarray}
 \Gamma(\tilde\ell \to\ell \chi_{1}^0+\ell \chi_{2}^0)& =&C\ 2
( m_Z\frac{  s_W^2}{M_1}-m_Z\frac{c_W^2}{M_2})^2, 
\label{eq:slep_chi120}\\
 \Gamma(\tilde\nu_\ell \to\nu_\ell \chi_{1}^0+\nu_\ell \chi_{2}^0)&=& C\ 2  
( m_Z\frac{  s_W^2}{M_1}+m_Z\frac{c_W^2}{M_2})^2,
\label{eq:snu_chi120}
\end{eqnarray}
with no dependence on $\tan\beta$.   Decays to charginos, however, show a different $\tan\beta$ dependence for $\tilde\ell$ and $\tilde\nu_\ell$.  The decay $\tilde\ell \rightarrow \nu_\ell \chi_1^\pm$ depends on $(c_\beta+ s_\beta \frac{\mu}{M_2})^2$, which decreases with increasing $\tan\beta$ until $\tan\beta \sim |M_2/\mu|$, when the decay branching fraction stabilizes.  On the other hand,
$\Gamma(\tilde\nu_\ell \rightarrow \ell \chi_1^\pm)$ depends only weakly on $\tan\beta$, since $c_\beta \mu/M_2$ is always small compared to $s_\beta$, which changes little for large $\tan\beta$.   As a result, the branching fractions for $\tilde\ell_L$ show a strong $\tan\beta$ dependence since the total decay width varies with $\tan\beta$ because of $\tilde\ell \rightarrow \nu_\ell \chi_1^\pm$, while the branching fractions for $\tilde\nu_\ell$ vary little with respect to $\tan\beta$.

\begin{figure}[h]
 \minigraph{7.8cm}{-0.2in}{(a)}{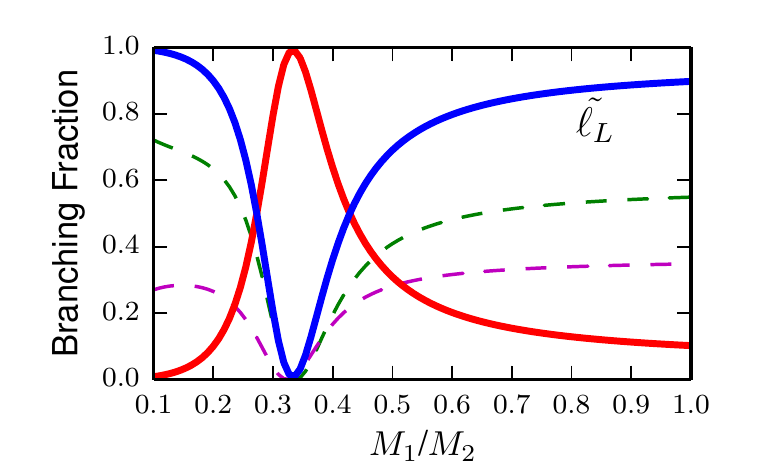}
\minigraph{7.8cm}{-0.2in}{(b)}{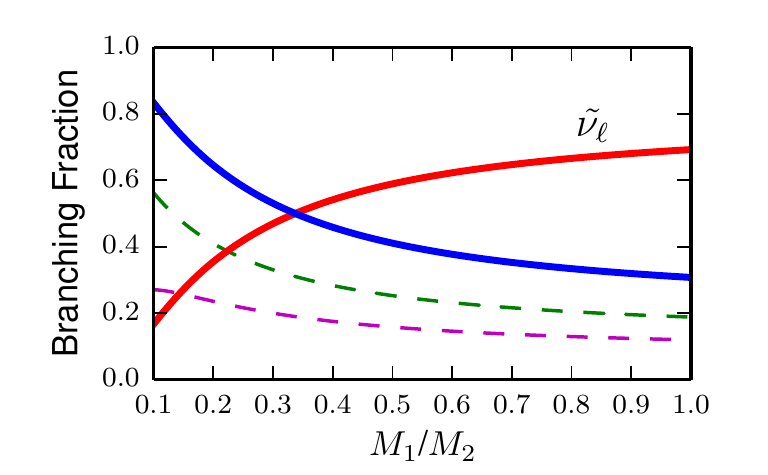}
\minigraph{7.8cm}{-0.2in}{(c)}{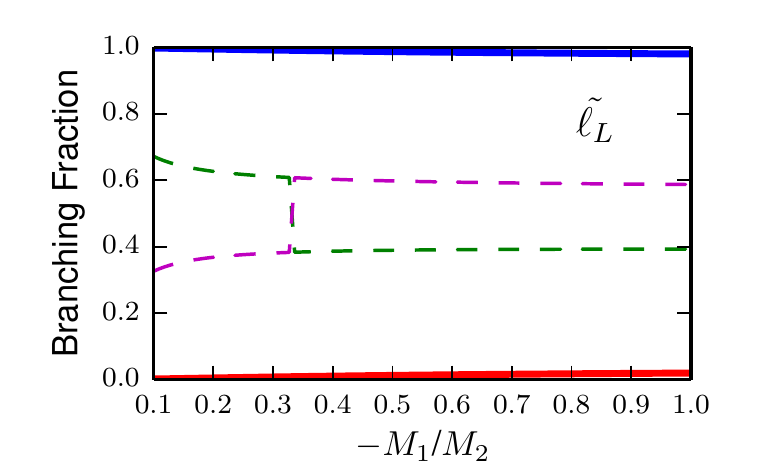}
\minigraph{7.8cm}{-0.2in}{(d)}{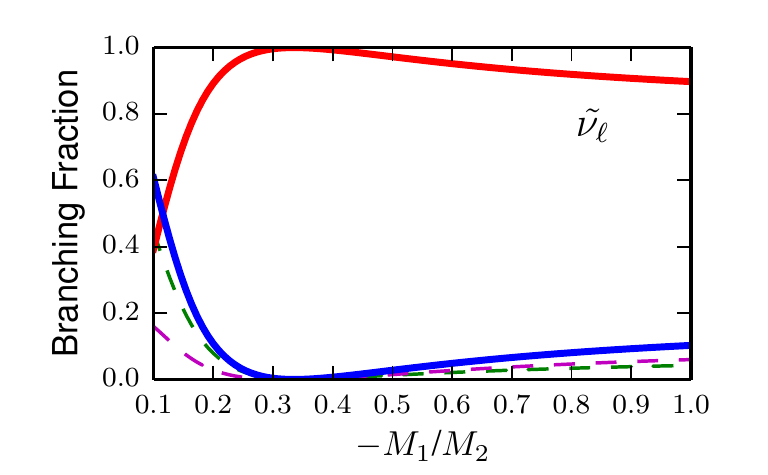}
  \caption{Branching fractions for  
$\tilde\ell_L \rightarrow \ell \chi_{1,2}^0, \nu \chi_{1}^\pm$ (left panels)    and $\tilde\nu \rightarrow \nu \chi_{1,2}^0, \ell\chi_{1}^\pm$   (right panels)  as a function of $M_1/M_2$. We have fixed $M_1/M_2>0$ in (a) and (b) and   $M_1/M_2<0$ in (c) and (d).  Other parameters are chosen as $m_{SL}=500$ GeV,  $|M_2|=$ 10 TeV, $\mu=100$ GeV and $\tan{\beta}=10$. The thick red and blue curves are the branching fractions to charginos and neutralinos ($\chi_1^0 + \chi_2^0$), respectively.  Also shown in dashed green and magenta are the individual decay branching fraction to $\chi_1^0$ and $\chi_2^0$.
 }
\label{fig:BR_slL_HIggsinoLSP_TB10}
\end{figure}

In Fig.~\ref{fig:BR_slL_HIggsinoLSP_TB10}, we show the branching fractions for charged slepton and sneutrino decays into Higgsino-like LSP $\chi_1^0$, as well as nearly degenerate Higgsino neutralino $\chi_2^0$ and chargino $\chi_1^\pm$.
 Other parameters are chosen to be $\tan\beta=10$, $m_{SL}=500$ GeV, $\mu=100$ GeV and $|M_2|= 10$ TeV.   In this paper, we always use $M_1>0$ as our convention. In general, there exist only two physical phases involving the electroweak gaugino and Higgsino mass parameters. We assume the gaugino/Higgsino sector introduces no new CP-violation, so these phases simply amount to relative signs. We chose them to be the relative signs of $M_1$ and $M_2$ and the relative sign of $\mu$ and $M_2$. As we discuss below, the choice of these phases can have a significant impact on the slepton decay branching fractions. On the other hand, 
the dependence of the branching fractions on the charged slepton/sneutrino mass or the Higgsino-like LSP mass is weak since the Higgsino-like neutralinos and charginos are almost degenerate and phase space effects cancel out.  Note that, within this Higgsino-like LSP regime, when $M_1$ or $M_2$ is less than $m_{\tilde{\ell}_L}$ and $m_{\tilde{\nu}_\ell}$,  the $\tilde{\ell}_L$ or $\tilde{\nu}_\ell$ first decay into the Bino or Wino-like states that subsequently cascade decay down to the Higgsino LSP.  The collider signature would be very different for such a case, which lies beyond the scope of the current study.

Fig.~\ref{fig:BR_slL_HIggsinoLSP_TB10} (a) shows the $M_1/M_2$ dependence of branching fractions   for $\tilde\ell_L$ to $\ell \chi_1^0$ (dashed green curve),  $\ell \chi_2^0$ (dashed magenta curve), as well as $\nu_\ell \chi_1^\pm$ (thick red curve), for $M_1/M_2>0$.  The sum of the $\ell \chi_1^0$ and $\ell \chi_2^0$ branching fractions is also given by the thick blue line since these two final states can not be distinguished at the LHC.  The curves show the limiting behavior for $M_1 \ll M_2$ where the decays are  dominated by the Bino component; for $M_1 \gtrsim M_2$ where the decays are dominated by the Wino component;  and   behavior in between.    For $M_1\ll M_2$, the branching fractions for decays to neutralinos reach almost 100\%  since the decay to charginos is suppressed by the relatively small  Wino fraction in $\chi_1^\pm$.  For  $M_1\gtrsim M_2$, the branching fraction for decays to neutralinos is about 90\%  since the decay to $\nu_\ell \chi_1^\pm$ is suppressed  by either $\cos\beta$ or $\mu/M_2$ compared to decay to neutralinos, as given in Eq.~(\ref{eq:slL_chi1pm}).

There is a  notable point at $M_1/M_2\sim \tan^2\theta_W\approx 0.3$ where the decays to neutralinos vanish due to the cancellation between the contributions of the Bino and Wino fractions in the Higgsino-like neutralinos for $M_1/M_2>0$.     In this region, decay to charginos, being all that remains, is dominant.

The branching fractions for $\tilde\nu_\ell$ decay are shown in Fig.~\ref{fig:BR_slL_HIggsinoLSP_TB10} (b).   For sneutrino decays to the chargino, the Wino-Higgsino mixing scales with $\sin{\beta}$ and so is generically more important than that for the charged slepton decays, unless the Bino component in $\chi_{1,2}^0$ dominates for small $M_1/M_2$.   No minimum for the decays to $\chi_{1,2}^0$ occurs since there is no cancellation between the Bino- and Wino- contribution for $M_1/M_2>0$.  Decay to neutralinos is dominant for $M_1 \ll M_2$, reaching about  80\% for $M_1/M_2=0.1$,  while decays to charginos dominate for $M_1 \gtrsim M_2$, reaching about 70\% for $M_1/M_2=1$.  

Fig.~\ref{fig:BR_slL_HIggsinoLSP_TB10} (c) and (d) show the the decays of the charged slepton and sneutrino for $M_1/M_2<0$.  
The $\tilde\ell_L\to \chi_{1,2}^0$ branching fraction will not have a minimum in its decay branching fraction since the Bino- and Wino-component interfere constructively in this case.     The step in the neutralino branching fraction curves near $M_1/M_2\sim0.3$ is due to a switchover between $\frac{1}{\sqrt{2}}(\tilde{H}_u\pm \tilde{H}_d)$ as being the LSP.   The branching fractions for $\tilde\ell_L \rightarrow \ell \chi_{1,2}^0$ almost reaches 100\%, due to the relative smallness of the partial decay width for $\tilde\ell_L \rightarrow \nu_\ell \chi_{1}^\pm$.   $\Gamma(\tilde\nu_\ell \rightarrow \nu_\ell \chi_{1,2}^0$), on the other hand, will experience a suppression for $M_1/M_2 \sim -\tan^2\theta_W$, as shown in Eq.~(\ref{eq:snu_chi120}).

\begin{figure}[h]
\minigraph{7.8cm}{-0.2in}{(a)}{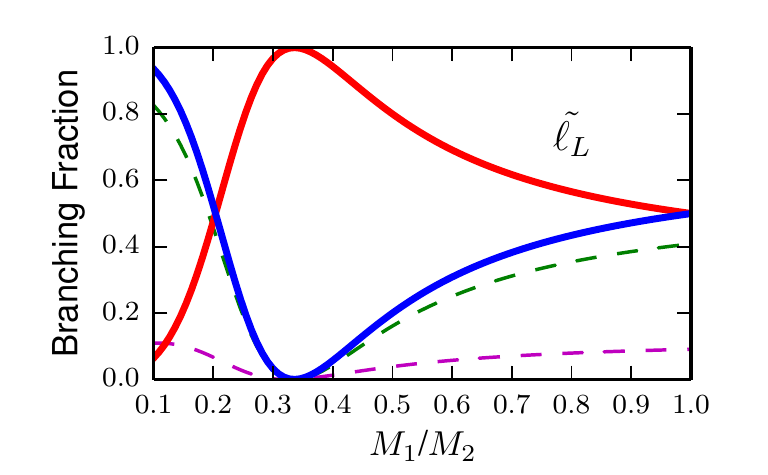}
\minigraph{7.8cm}{-0.2in}{(b)}{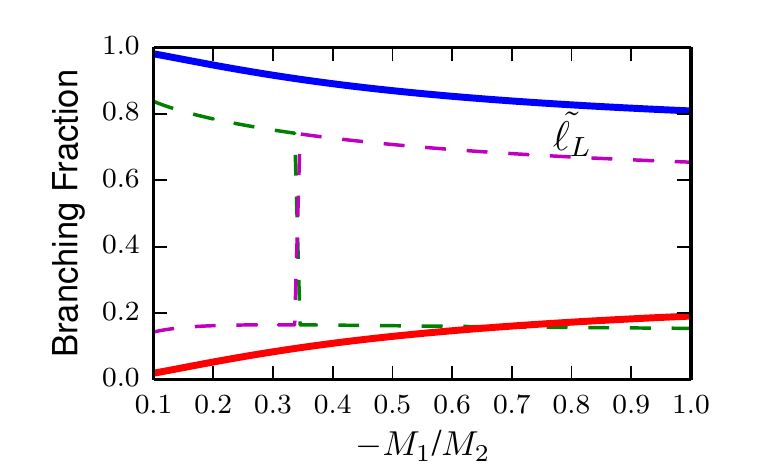}
  \caption{Branching fraction of 
$\tilde\ell_L \rightarrow \ell \chi_{1,2}^0, \nu \chi_{1}^\pm$ 
 as a function of $M_1/M_2$ for (a) $M_1/M_2>0$ and (b) $M_1/M_2<0$ with $\tan\beta=3$.    The other parameter choices and color coding are the same as in Fig.~\ref{fig:BR_slL_HIggsinoLSP_TB10}.}
\label{fig:BR_slL_HIggsinoLSP_TB3}
\end{figure}

Fig.~\ref{fig:BR_slL_HIggsinoLSP_TB3}  shows the dependence of charged slepton branching fractions on $M_1/M_2$ for $\tan\beta=3$.   While the generic features are the same as Fig.~\ref{fig:BR_slL_HIggsinoLSP_TB10} for $\tan\beta=10$, the decay fraction for $\tilde\ell \rightarrow \nu_\ell \chi_1^\pm$ is relatively larger due to the enhancement of $\Gamma(\tilde\ell \rightarrow \nu_\ell \chi_1^\pm)$ arising from the larger value of $\cos\beta$.   For $M_1/M_2=1$, decay branching fractions to $\ell \chi_{1,2}^0$ and  $\nu_\ell \chi_{1,2}^0$ are about 50\% each.  Similarly, for $M_1/M_2<0$,  the branching fraction of decays to neutralinos is about  80\% to 100\%, while the decays to charginos could be as large as 20\%.

\begin{figure}[h]
 \minigraph{7.8cm}{-0.2in}{(a)}{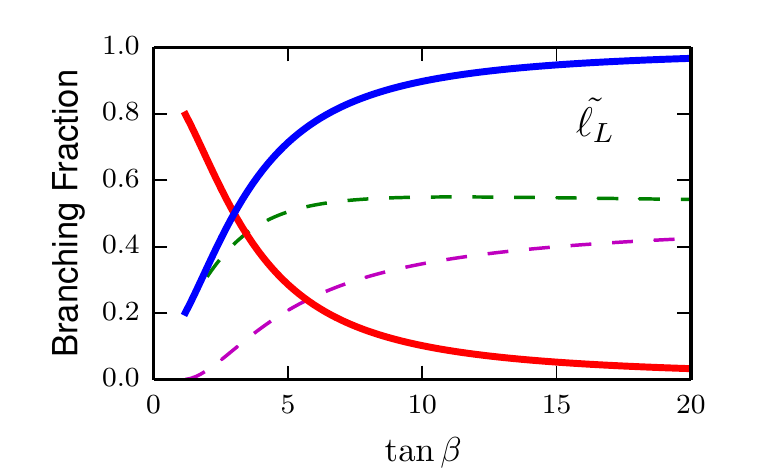}
\minigraph{7.8cm}{-0.2in}{(b)}{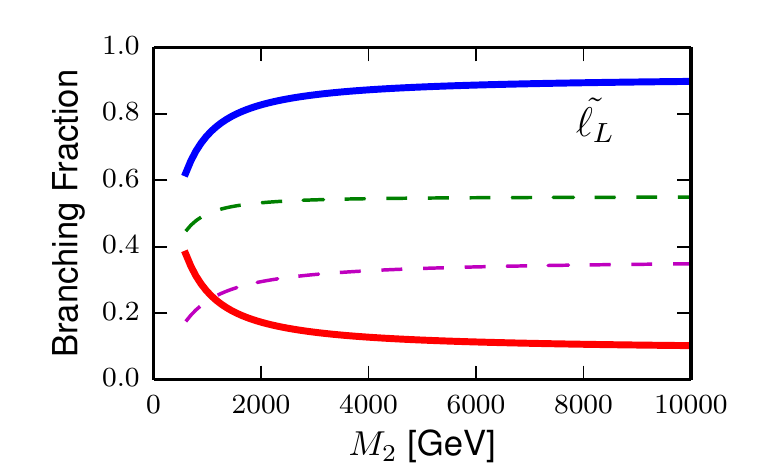}
   \caption{Branching fraction of  $\tilde\ell_L \rightarrow \ell \chi_{1,2}^0, \nu \chi_{1}^\pm$ as a function of (a) $\tan\beta$   for $M_2=10$ TeV and (b) $M_2$ for $\tan\beta=10$.  We have chosen the other parameters to be $m_{SL}=500$ GeV, $\mu=100$ GeV,  $M_1/M_2=1$.  }
\label{fig:BR_slL_HIggsinoLSP_beta}
\end{figure}

Fig.~\ref{fig:BR_slL_HIggsinoLSP_beta} (a)  shows the $\tan\beta$ dependence for the $\tilde\ell_L \rightarrow \ell \chi_{1,2}^0, \nu \chi_{1}^\pm$ branching fractions for $m_{SL}=500$ GeV, $\mu=100$ GeV,  $M_1/M_2=1$ and $M_2=10$ TeV. 
For $\tan\beta < M_2/\mu$ such that  $\cos\beta$ is much greater than $s_\beta \mu/M_2$, the $\tilde\ell_L \rightarrow \nu_\ell \chi_1^\pm$ branching fraction always decreases as $\tan\beta$ increases, with $\tilde\ell_L \rightarrow \ell \chi_{1,2}^0$ becoming dominant for $\tan\beta\gtrsim10$.    Fig.~\ref{fig:BR_slL_HIggsinoLSP_beta} (b)  shows the $M_2$ dependence of charged slepton decay branching fraction for $\tan\beta=10$.  The dependence of charged slepton decay branching fractions on $M_2$ is also weak unless $\tan\beta > M_2/\mu$, when $\Gamma(\tilde\ell \to\nu_\ell  \chi_{1}^\pm)$ could have an explicit $M_2$ dependence.    The  $\tilde\ell_L \rightarrow \nu_\ell \chi_1^\pm$ branching fraction decreases as $M_2$ increases, saturating when $ M_2/\mu > \tan\beta$.     The sneutrino decay branching fraction, on the other hand, depends mildly on $\tan\beta$ and $M_2$. 

Note that in the foregoing discussion of the $\tilde\ell_L$ and $\tilde\nu_\ell$ decays to Higgsino-like LSPs, we have considered the case of $M_1>0$ and $\mu>0$, with two different signs for $M_2$.  The relative sign between these three mass parameters is physical, and the behavior of the branching fractions will change when one of these parameters flips sign.  For $\mu/M_2<0$,  decays to charginos will be relatively suppressed compared to the $\mu/M_2>0$ case, in particular for $\tilde\ell_L$, as shown in Eq.~(\ref{eq:slL_chi1pm}).

For the  $\tilde{\ell}_R$, it again decays to $\ell \chi_{1,2}^0$ 100\% via the Bino-component of $\chi_{1,2}^0$ since the decay to $\chi_1^\pm$   is suppressed by the small lepton Yukawa couplings.

\begin{table}
\begin{center}
\begin{tabular}{|l|c|c|c|c|c|}
\hline 
& $\tilde\ell_L \rightarrow \ell \chi_{1(2)}^0$&$\tilde\ell_L \rightarrow \nu \chi_1^\pm$&$\tilde\nu \rightarrow \nu \chi_{1(2)}^0$&$\tilde\nu \rightarrow \ell \chi_1^\pm$&$\tilde\ell_R \rightarrow \ell \chi_{1(2)}^0$ \\ \hline
Bino-like LSP&100\%&&100\%&&100\%\\
Wino-like LSP&33.3\%&66.7\%&33.3\%&66.7\%&100\% \\
  Higgsino-like LSP (I)& 0.8\%& 99.2\%& 50.3\%& 49.7\%&100\% \\
  Higgsino-like LSP (II)& 99.1\%& 0.9\%& 0.0\%& 100.0\%&100\% \\
 \hline
\end{tabular}
\end{center}
\caption{Branching fractions of charged sleptons and sneutrinos into Bino-, Wino- and Higgsino-like LSPs.  We have set $m_{SL} = 500$ GeV,  $\tan\beta=10$, and used an LSP mass parameter of 100 GeV.   For the Higgsino-like LSP case, we  presented the results for two representative benchmark values: (I) $M_1/M_2=1/3$ and (II) $M_1/M_2=-1/3$ with $|M_2|=10$ TeV.    }
\label{tab:BR} 
\end{table}

Given the branching fraction dependence on $M_1/M_2$, as well as $\tan\beta$, for the Higgsino-like LSP case, we consider two benchmark choices for $M_1/M_2$ to  represent two extreme cases: (I) $M_1/M_2=1/3$ with suppressed $\Gamma(\tilde\ell_L \rightarrow \ell \chi_{1,2}^0)$ and (II) $M_1/M_2=-1/3$ with suppressed $\Gamma(\tilde\nu_\ell \rightarrow \nu_\ell  \chi_{1,2}^0)$ (therefore enhanced decays to charged leptons).    The corresponding branching fractions are given in Table.~\ref{tab:BR}.   Case (I) leads to a suppressed overall cross section for dilepton plus $\met$ final states, while case (II) leads to an enhancement. These cases are the upper and lower boundaries of the envelope of possible signal rates in the Higgsino-like LSP scenario.

\subsection{Slepton production and signatures}

For Drell-Yan pair production of sleptons $\tilde{\ell}_L\tilde\ell_L$, $\tilde{\ell}_L\tilde\nu_\ell$, $\tilde{\nu}_\ell\tilde\nu_\ell$ and $\tilde{\ell}_R\tilde\ell_R$ with dominant direct decay of sleptons into $\chi_1^0$ (and $\chi_1^\pm$, $\chi_2^0$ for Wino-like and Higgsino-like LSP cases), the collider signatures are dilepton plus $\met$, single lepton plus $\met$, and $\met$ only.  The single lepton channel suffers from large SM backgrounds, mainly driven by $W$ boson production.  The $\met$  only signature requires an extra jet or lepton from initial or final state radiation, which leads to more suppressed cross sections.  Current collider searches for slepton Drell-Yan production focus on the final state of two isolated energetic leptons plus $\met$ \cite{Aad:2014vma, Khachatryan:2014qwa}.  The SM backgrounds are typically large, dominantly from $WW$ or $t\bar{t}$.  
 In our analyses below, we focus on the dilepton plus $\met$  channel and reinterpret the  current   8 TeV  LHC slepton search limits  for various LSP scenarios, as well as project the reach of the LHC at 14 TeV. In particular, we include contributions from the presence of sneutrinos for the case of left-handed sleptons, as their mass is related to the left-handed slepton mass and they can contribute to the dilepton and missing energy signature for non-Bino-like LSPs.

\begin{figure}[h]
\minigraph{7.8cm}{-0.2in}{(a)}{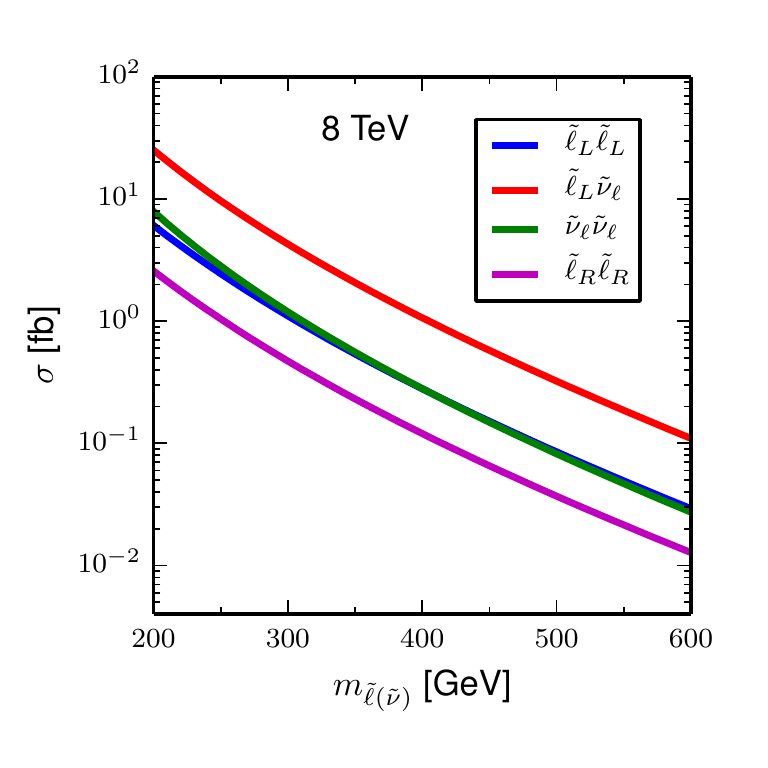}
\minigraph{7.8cm}{-0.2in}{(b)}{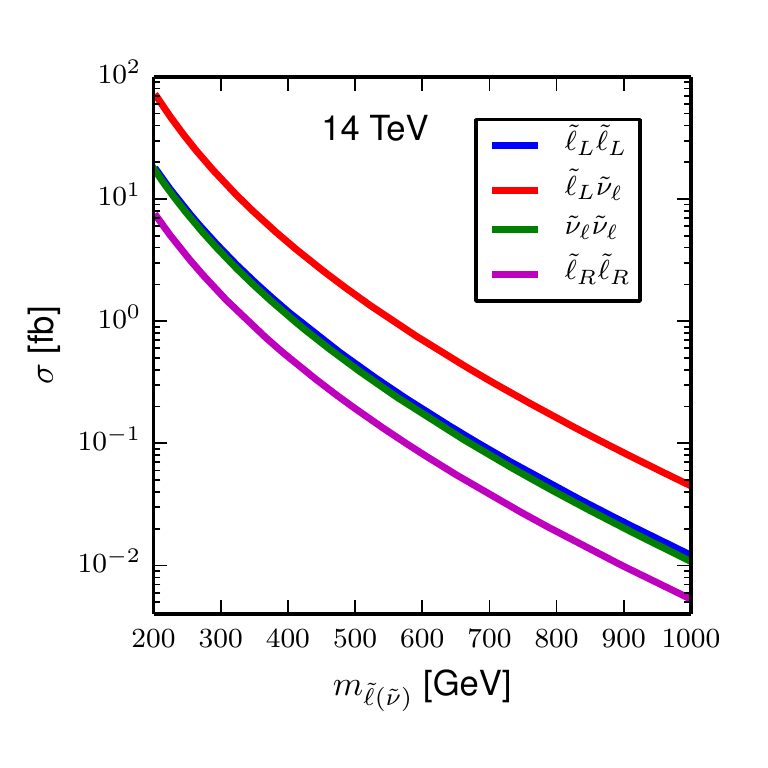}
 \caption{Leading-order cross sections for the Drell-Yan pair production of $\tilde\ell_L \tilde\ell_L$, $\tilde\ell_L\tilde\nu_\ell$, $\tilde\nu_\ell \tilde\nu_\ell$ and $\tilde\ell_R \tilde\ell_R$ for the (a) 8 TeV and (b) 14 TeV LHC. }
\label{fig:CS}
\end{figure}

In Fig.~\ref{fig:CS}, we show the individual leading order Drell-Yan pair production cross sections for $\tilde\ell_L \tilde\ell_L$, $\tilde\ell_L \tilde\nu_\ell$, $\tilde\nu_\ell \tilde\nu_\ell$ and $\tilde\ell_R \tilde\ell_R$ at the LHC with $\sqrt{s}=8$ TeV and 14 TeV.  The  production of $\tilde\ell_L\tilde\nu_\ell$ is markedly larger than the other cross sections, ranging from 25 to 0.1 fb for $\sqrt{s}=8$ TeV for masses from 200 to 600 GeV and from 70 to 0.04 fb for $\sqrt{s}=14$ TeV for masses from 200 GeV to 1 TeV. Sneutrino and left-handed charged slepton pair production are comparable in size and smaller than the  production by about a factor of  3, while right-handed slepton pair production, due to a cancellation between the Z and $\gamma$ $s$-channel graphs, is smaller still, about an order of magnitude less than the associated production cross section. The NLO K-factors are approximately 1.18 \cite{Fuks:2013lya}, independent of which particular pair production considered, as the QCD structure of the graphs is identical in all four cases.

\begin{figure}[h]
\minigraph{7.8cm}{-0.4in}{(a)}{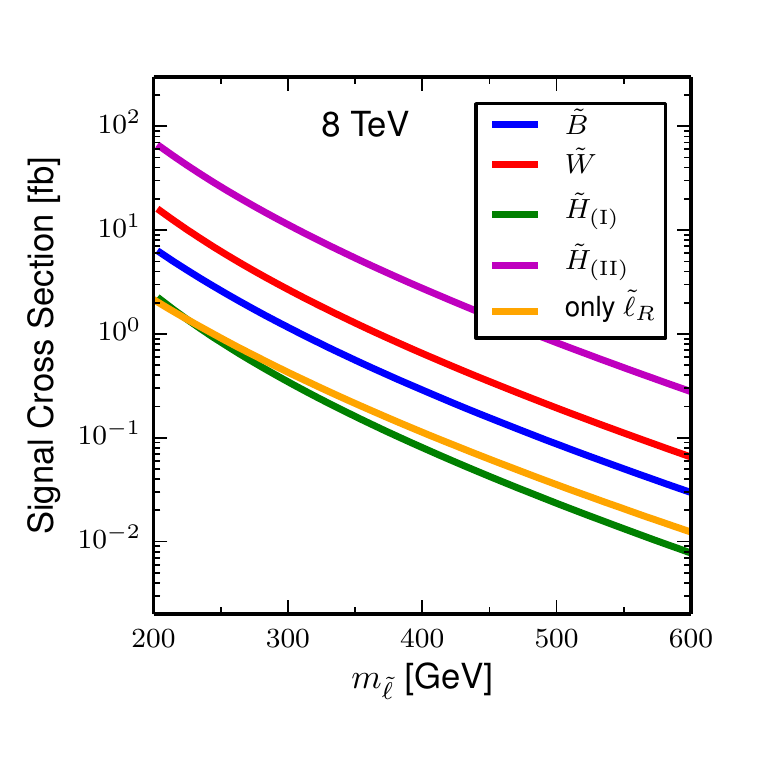}
\minigraph{7.8cm}{-0.4in}{(b)}{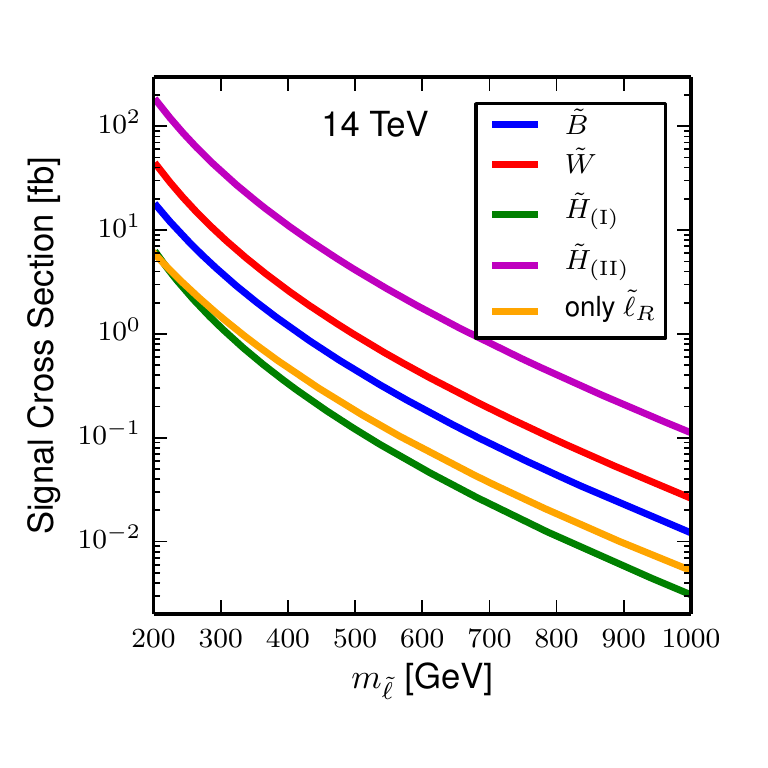}
\caption{Sum of all possible production mechanisms weighted by branching fraction for dilepton plus $\met$ final states as a function of slepton mass for the (a) 8 TeV  and (b) 14 TeV   LHC.  
 }
\label{fig:CSBR}
\end{figure}

In Fig.~\ref{fig:CSBR}, we show the signal cross section,
the sum of all possible slepton production cross sections multiplied by the branching fraction 
leading to dilepton plus $\met$ final states as a function of slepton mass for the (a) 8 TeV  and (b) 14 TeV  LHC.   For the left-handed slepton, we have included the contributions from $\tilde\ell_L\tilde\ell_L$,  $\tilde\nu_L\tilde\nu_L$ and $\tilde\ell_L\tilde\nu_L$.  The Higgsino-like LSP benchmark (II) for light $\tilde\ell_L$ with $M_1/M_2= -1/3$ represents the most promising scenario since sleptons decay dominantly to $\ell \chi_{1,2}^0$ while sneutrinos decay dominantly to $\ell \chi_{1}^\pm$.  The cross sections range from   about 70 fb to 0.3 fb  for slepton masses  from 200 to 600 GeV at the 8 TeV LHC, and from 200 fb to 0.1 fb  at the 14 TeV LHC for slepton masses from 200 GeV to 1 TeV.  The Higgsino-like LSP benchmark (I) for light $\tilde\ell_L$ with $M_1/M_2= 1/3$  represents the worst-case scenario with a strong suppression of slepton decays to leptons. For the Bino- and Wino-like LSP scenarios, the signal cross sections 
range from 40 fb to about 0.01 fb for slepton mass between 200 GeV  to 1 TeV at the 14 TeV LHC. Right-handed sleptons are less promising than all the left-handed cases except for the highly pessimistic Higgsino benchmark (I), ranging in signal cross section from 7 fb to 0.005 fb at the 14 TeV LHC.

\begin{figure}[h]
\minigraph{7.8cm}{-0.2in}{(a)}{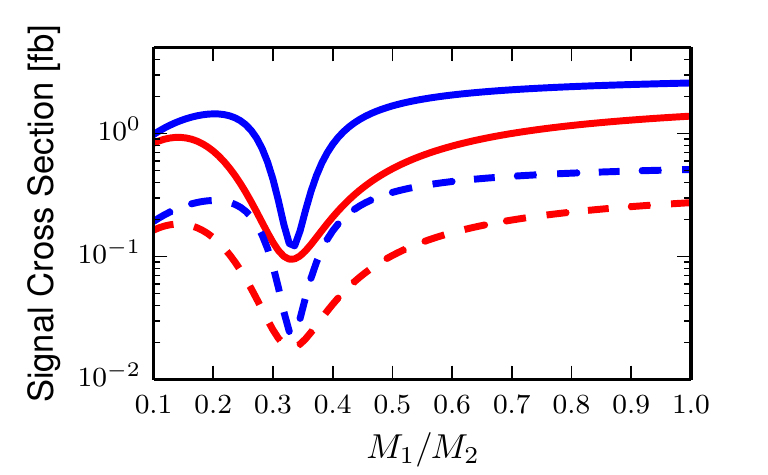}
\minigraph{7.8cm}{-0.2in}{(b)}{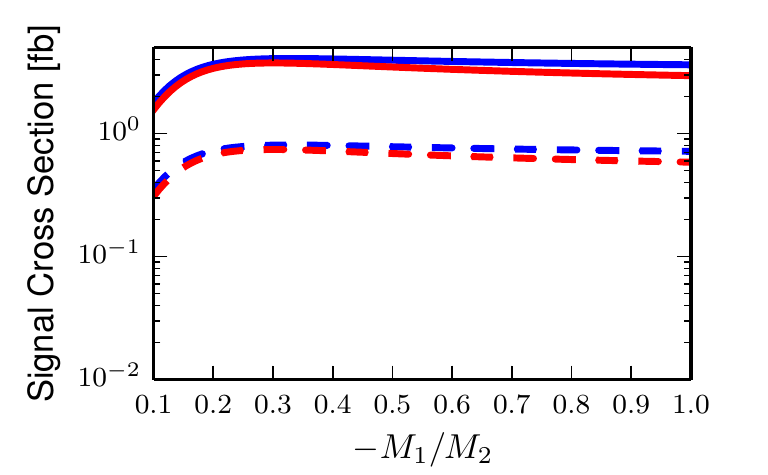}
\caption{ $\sigma\times{\rm Br}$ for dilepton plus $\met$ final states in a Higgsino-like LSP scenario as a function of $M_1/M_2$ for (a) $M_1/M_2>0$ and (b)  $M_1/M_2<0$.  Solid curves and dashed curves are for 14 TeV and 8 TeV, respectively.  Blue curves are for $\tan\beta=10$ and Red curves are for $\tan\beta=3$.  Other parameters are chosen as $m_{SL}=500$ GeV and $|M_2|=10$ TeV.
 }
\label{fig:CSBR_M1overM2}
\end{figure}

To show the strong dependence of  $\tilde{\ell}_L$, $\tilde\nu$ decay  branching fractions on the sign and value of $M_1/M_2$, in Fig.~\ref{fig:CSBR_M1overM2} we plot the $\sigma\times{\rm Br}$ for dilepton plus $\met$ final states for the Higgsino-like LSP case with a light $\tilde\ell_L$   as a function of $M_1/M_2$ at the 14 (8) TeV LHC, which are indicated by solid (dashed) curves. The dip in the positive $M_1/M_2$ case results from the suppressed charged slepton decay branching fractions to leptons at $M_1/M_2 \sim \tan^2\theta_W$,  with an overall cross section of about 0.1 fb at the 14 TeV LHC with $\tan\beta=10$.
The maximum value for $\sigma\times{\rm Br}$ appears at $M_1/M_2 \sim -\tan^2\theta_W$ due to the enhanced sneutrino decay branching fractions to leptons, with an overall cross section of about 4.0 fb.  For $|M_1/M_2| \ll 1$, the Bino component in the Higgsino states $\chi_{1,2}^0$ is dominant and the cross section is about 1 fb at the 14 TeV LHC, while the cross section reaches about  2.6 $-$ 3.6 fb for $|M_1/M_2| \gtrsim 1$. Smaller values of $\tan\beta$ typically lead to smaller signal cross sections.

\section{Current  searches and studies}
\label{sec:limit}

The   least model-dependent bounds on sleptons  are obtained from LEP searches for dilepton plus missing energy signatures \cite{LEPslepton}
with $\sqrt{s}$ up to 208 GeV.  For a  slepton-neutralino LSP mass splitting greater than 15 GeV, the  right-handed slepton mass limits are: $m_{\tilde{e}_R} > 99.6$
GeV, $m_{\tilde{\mu}_R} > 94.9$ GeV and $m_{\tilde{\tau}_R} > 85.9$ GeV.
For left-handed sleptons  with a Bino-like LSP, the bounds are stronger due to the larger production cross section.
  For tau sleptons, on the other hand,  the presence of significant left-right mixing can decrease the
production cross section for the lightest stau pair, leading to more relaxed limits. 
A lower limit of
$m_{\tilde{\tau}} > 85.0$ GeV can be obtained when the production
cross section for the lightest stau is minimized.  It should be noted
that the slepton mass limits are obtained with $\mu=-200$ GeV and
$\tan\beta=1.5$, a point at which the neutralino mass limit based on
the LEP neutralino and chargino searches is the weakest, and the
selectron cross section is relatively small.  

The foregoing bounds also assume the gaugino mass unification
relation $M_1=(5/3)\tan^2\theta_W M_2$, which is relevant in
fixing the neutralino mass and field content, with the neutralino LSP being mostly Bino-like.  Slepton mass
limits would change for a non-unified mass relation between $M_1$ and
$M_2$.
 In the case where the $\tilde{e}_R - \chi_1^0$ mass splitting is small and
the usual   dilepton   search is insensitive, a single lepton plus
missing energy search yields a lower limit on $m_{\tilde{e}_R}$ of 73
GeV, independent of
$m_{\chi_1^0}$ \cite{LEPsinglelep}.  For sneutrinos, a mass limit of
45 GeV can be deduced from the invisible $Z$ decay width \cite{:2005ema}.  
An indirect mass limit on sneutrinos can also be derived
from the direct search limits on the charged slepton masses, but for LEP searches it is not competitive with the invisible width constraint.

Searches for first and second generation charged sleptons have been performed by both the ATLAS \cite{sleptonATLAS,Aad:2014vma} and CMS collaborations \cite{sleptonCMS}.  With about 20 ${\rm fb}^{-1}$ luminosity collected at 8 TeV, both collaborations studied the signal of  opposite-sign (OS) same flavor (SF) dilepton plus missing $E_T$ from the electroweak pair production of sleptons assuming a 100\% decay branching fraction for  ${\tilde\ell}^\pm\to \ell^\pm +\chi_1^0$.   
The most stringent bounds  come from the ATLAS results, which exclude left-handed (right-handed) slepton masses between 95 and 310 GeV  (235 GeV) at 95\% C.L. for a Bino-like LSP with $m_{\chi_1^0}=0$ GeV.   For larger  $\chi^0_1$ masses, the upper range of the exclusion reach does not change while the lower bound shifts  approximately as 80 GeV $+m_{\chi_1^0}$. 
    
\section{ Recasting LHC 8 TeV Search Limits}
\label{sec:7TeV}
 
We consider the signal consisting of two same flavor, opposite sign energetic leptons (electrons or muons) plus significant  missing energy at the 8 TeV LHC.    The dominant SM backgrounds arise from $t{\bar t}$ and di-boson production.   We use Madgraph 5 version v1.4.7 and Madevent v5.1.4.7 \cite{Alwall:2011uj} to generate our signal events.  These events are passed to Pythia v6.426 \cite{Sjostrand:2006za}  to simulate initial state radiation, final state radiation, showering and hadronization.  Additionally we use Delphes v3.0.10 \cite{deFavereau:2013fsa} with the Snowmass card   \cite{snowmass} to simulate detector effects. We chose not to simulate pile-up to increase computational speed because we are considering a clean leptonic final state which should not be sensitive to pile-up.  The event generation procedure produces events at leading order.  NLO effects are taken into account by scaling our events by an appropriate K-factor \cite{Fuks:2013lya}.  We additionally take into account various experimental efficiencies that may be poorly modeled by our crude detector simulation by scaling our signal 
 yields to match the expected yields quoted in the experimental search \cite{sleptonATLAS}.

Following the 8 TeV dilepton search technique at the ATLAS \cite{sleptonATLAS,Aad:2014vma},  we apply the following cuts:
\begin{itemize}
\item Exactly two leptons (electron or muon) with $p_T^\ell>10$ GeV and $|\eta^\ell|<2.5$.  The invariant mass of the lepton pair is required to be greater than 20 GeV and  to be away from the $Z$-pole: $|m_{\ell\ell}-m_Z|>10$ GeV.

\item Jet veto with $p_{T}^j<20$ GeV for central jets with $|\eta^j|<2.4$;  $p_{T}^j<30$ GeV for forward jets with $2.4<|\eta^j|<4.5$. 

\item $\met^{{\rm rel}}>40$ GeV, with 
\begin{equation}
\met^{{\rm rel}}=\left\{
\begin{array}{ll}
\met\sin\left(\Delta\phi^{\ell, j}\right)&  {\rm for\ }  \Delta\phi^{\ell, j}<\pi/2 \\
\met & {\rm otherwise}
\end{array}
\right.  ,
\end{equation}
where  $\Delta\phi^{\ell, j}$ is the azimuthal angle between the direction of $p_T^{\rm miss}$ and the nearest lepton or central jet.

\item $M_{T2}>$ 90 or 110  GeV where $M_{T2}$ is the stransverse mass variable \cite{Lester:1999tx, Barr:2003rg, Cheng:2008hk}.   We choose the optimized cut to give the higher value of $S/\sqrt{B}$ for each point in signal parameter space, where $S$ ($B$) is the number of signal (background) events.
\end{itemize}

For the signal process, our simulation matches well with the ATLAS distributions for the given benchmark points after a scaling by factor of 1.25 for both di-electron and di-muon channels that accounts for both a K-factor expected to be 1.18 and differences in reconstruction efficiencies. We use this scaling factor in all subsequent calculations for 8 TeV.

\begin{figure}[h]
 \minigraph{7.8cm}{-0.2in}{(a)}{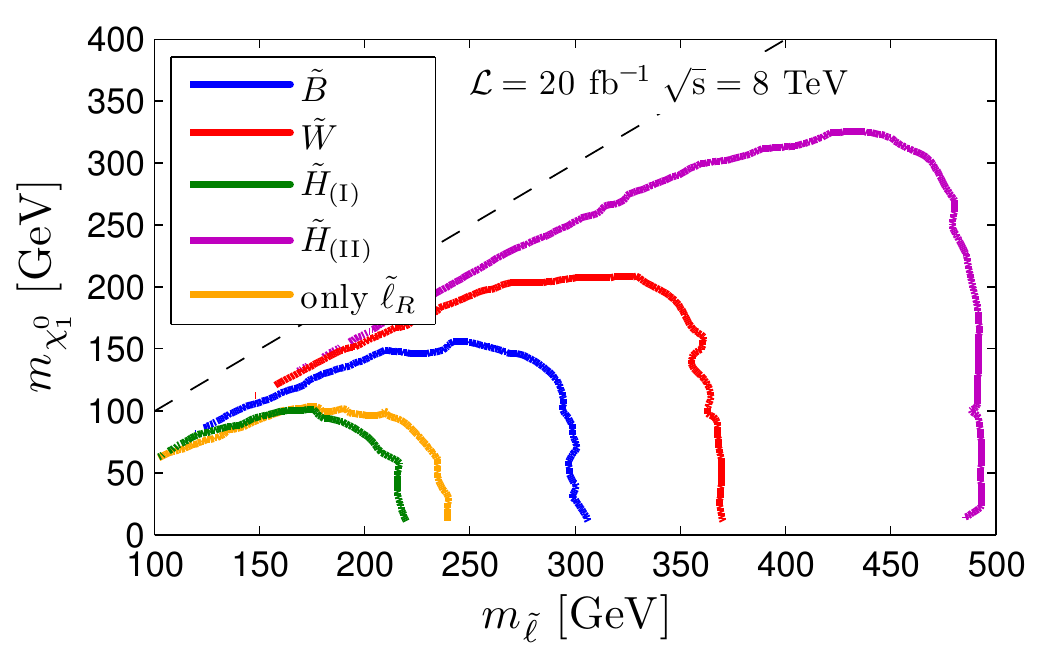}
\minigraph{7.8cm}{-0.2in}{(b)}{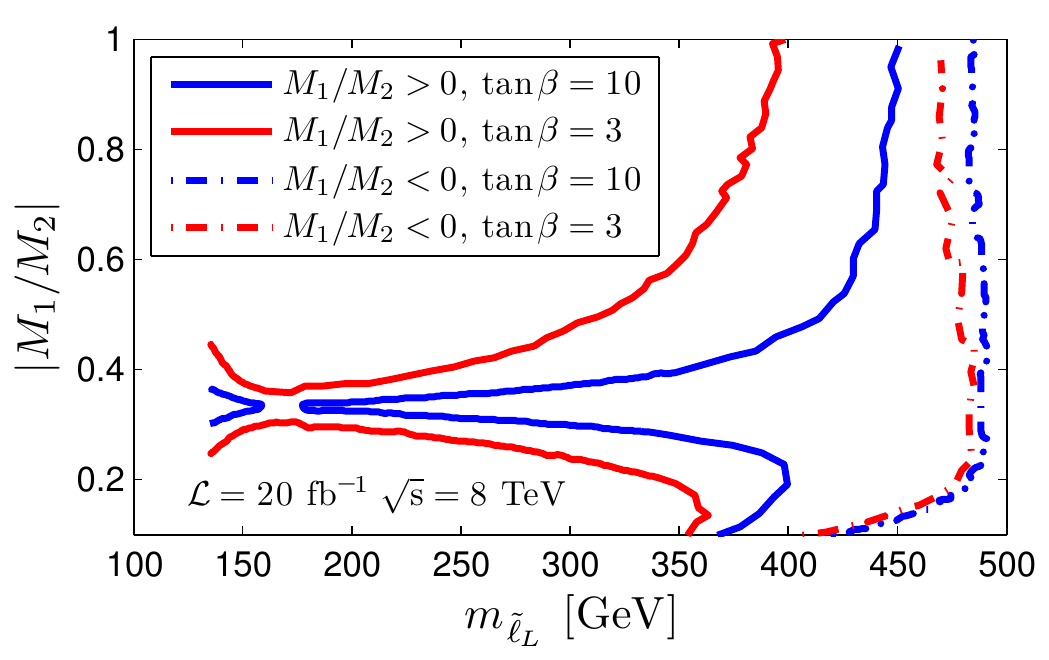}
  \caption{Recast of ATLAS dilepton plus $\met$ search results \cite{sleptonATLAS}  with 20 ${\rm fb}^{-1}$ luminosity data collected at the 8 TeV LHC. 
The left panel shows the 95\% C.L. exclusion limits in the $m_{\tilde\ell} - m_{\tilde\chi_1^0}$ plane for the left-handed slepton with Bino-like LSP (blue line), Wino-like LSP (red line), Higgsino-like LSP with $M_1/M_2=1/3$ (green line), $M_1/M_2=-1/3$ (magenta line), as well as for the right-handed slepton (orange line).      The right panel shows the 95\% C.L. exclusion bounds in the $m_{\tilde\ell_L} - |M_1/M_2|$ plane for the left-handed slepton with a Higgsino-like LSP, for $M_1/M_2>0$ (solid lines)  with $\tan\beta=10$ (blue), 3 (red), and $M_1/M_2<0$ (dash-dotted lines).    All other parameters are fixed to be $\mu=100$ GeV and $|M_2|=10$ TeV.   }
\label{fig:8TeV_recast}
\end{figure}

To reproduce the exclusion plot from the ATLAS paper we utilized the $CL_s$ method
discussed in \cite{Junk:1999kv, Read:2000ru}.  We generate signal events using Monte-Carlo to determine the signal strength over the range of parameters 
$90 < m_{\tilde\ell} < 600$ GeV and $25 < m_{\chi_1^0} < m_{\tilde\ell} - 30$ GeV.  
For our SM backgrounds, we simply use the number of events predicted by the ATLAS experiment for each cut scenario \cite{sleptonATLAS}.  
We follow the method in the ATLAS paper where we choose the $M_{T2}$ cut that maximizes $S/\sqrt{B}$. We also demand that the signal to background is greater than a minimal threshold:  $S/B>0.1$.   We reproduced the  exclusion limits for the Bino-like LSP for left- and right-handed sleptons, indicated by the solid blue and orange curves, respectively, in Fig. \ref{fig:8TeV_recast} (a).  Our bounds reproduce the ATLAS search results well with  slight discrepancy at low masses.  
  
As discussed in detail in Sec.~\ref{sec:MSSMslepton}, the decay branching fractions of left-handed charged sleptons and sneutrinos  depend strongly on the composition of the neutralino LSP, and in particular on the sign and value of $M_1/M_2$ in the Higgsino-like LSP scenario. For a given slepton mass, the resulting dilepton $+\met$ final states cross section, therefore, varies with the choice of LSP scenario. In Fig.~\ref{fig:8TeV_recast} (a), we recast the current 8 TeV ATLAS slepton search results in the dilepton $+\met$ channel for the various benchmark scenarios given in Table~\ref{tab:BR}. For cases where $m_{\tilde{\ell}}\gtrsim m_{\tilde{\chi}}+50$ GeV , we find that the Wino-like LSP scenario is excluded for slepton masses below approximately 365 GeV, while the pessimistic and optimistic Higgsino-like LSP scenarios imply exclusion of sleptons lighter  than about 220 GeV and 495 GeV, respectively. 

To show the dependence of  limits on $M_1/M_2$ in the Higgsino-like LSP scenario, we plot in Fig.~\ref{fig:8TeV_recast} (b) the 95\% C.L. limits in the parameter space of $|M_1/M_2|$  versus $m_{\tilde\ell_L}$    with  the Higgsino-like LSP mass set to be 100 GeV. The solid curves  are for $M_1/M_2>0$ whereas the dash-dotted curves are for $M_1/M_2<0$.   Regions to the left of the curves are excluded (excepting the small blue wedge of unconstrained light sleptons on the left edge of the plot). The suppression of signal for positive $M_1/M_2\sim\tan^2\theta_W$ is clearly visible in the blue curves, where sensitivity drops precipitously to much lower masses. We also note that the low mass region of $m_{\tilde\ell}\lesssim150$ GeV for $\tan\beta=10$ cannot be excluded for this LSP mass due to the loss of sensitivity for small mass splitting between the slepton and the LSP. The exclusion region for $\tan\beta=3$ is even weaker due to the suppression of the signal cross sections, with no sensitivity for any slepton masses  when $0.3 <  M_1/M_2 < 0.35$.    In the negative $M_1/M_2$ case, by comparison, the  slepton mass exclusion is  significantly stronger ($m_{\tilde\ell}\gtrsim  470 - 490$ GeV) and relatively    insensitive to  $|M_1/M_2|$ until it gets fairly small, when the $m_{\tilde\ell}$ reach is reduced due to the suppression of $\Gamma(\tilde\nu_L \rightarrow \ell \chi_1^\pm)$.

\section{14 TeV Exclusion and Discovery Reach} 
\label{sec:collider}

We now turn to projections for Run II at the LHC, with 14 TeV center-of-mass energy. As with the 8 TeV LHC analysis, we consider the dilepton $+\met$ channel and generate the signal and background Monte-Carlo events in the same manner as in Sec.~ \ref{sec:7TeV}. For the signal process, we used the next-to-leading order production cross section for sleptons as given in Ref.~\cite{Fuks:2013lya}.  Background processes are scaled with K-factors from Ref.~\cite{Bern:2008ef}. We generate the signal over the range of parameters  $200 < m_{\tilde\ell} < 1000$ GeV and $25 < m_{\chi_1^0} < m_{\tilde\ell} - 30$ GeV. We also demand $S>2$, $B>2$,   and $S/B > 0.1$.   

For the 14 TeV analysis, we adopted the following cuts: 
\begin{itemize}
\item 2 isolated leptons (electron or muon) with $p_T^\ell >50$ GeV, $|\eta^\ell|<2.5$, and $m_{\ell\ell} > 20$ GeV.
\item No jets with $p_T^j>50$ GeV and $|\eta^j|<4.5$.
 \item $Z$ veto with $|m_{\ell\ell}-m_Z|>10$ GeV. 
 \item Optimized cuts on  $\met^{{\rm rel}}$ and $M_{T2}$. Cuts range from $100 < \met^{{\rm rel}} < 200$ GeV and $0 < M_{T2} < 200 $ GeV in increments of 50 GeV. 

\end{itemize}

\begin{figure}[h]
\begin{center}
 \minigraph{7.8cm}{-0.2in}{(a)}{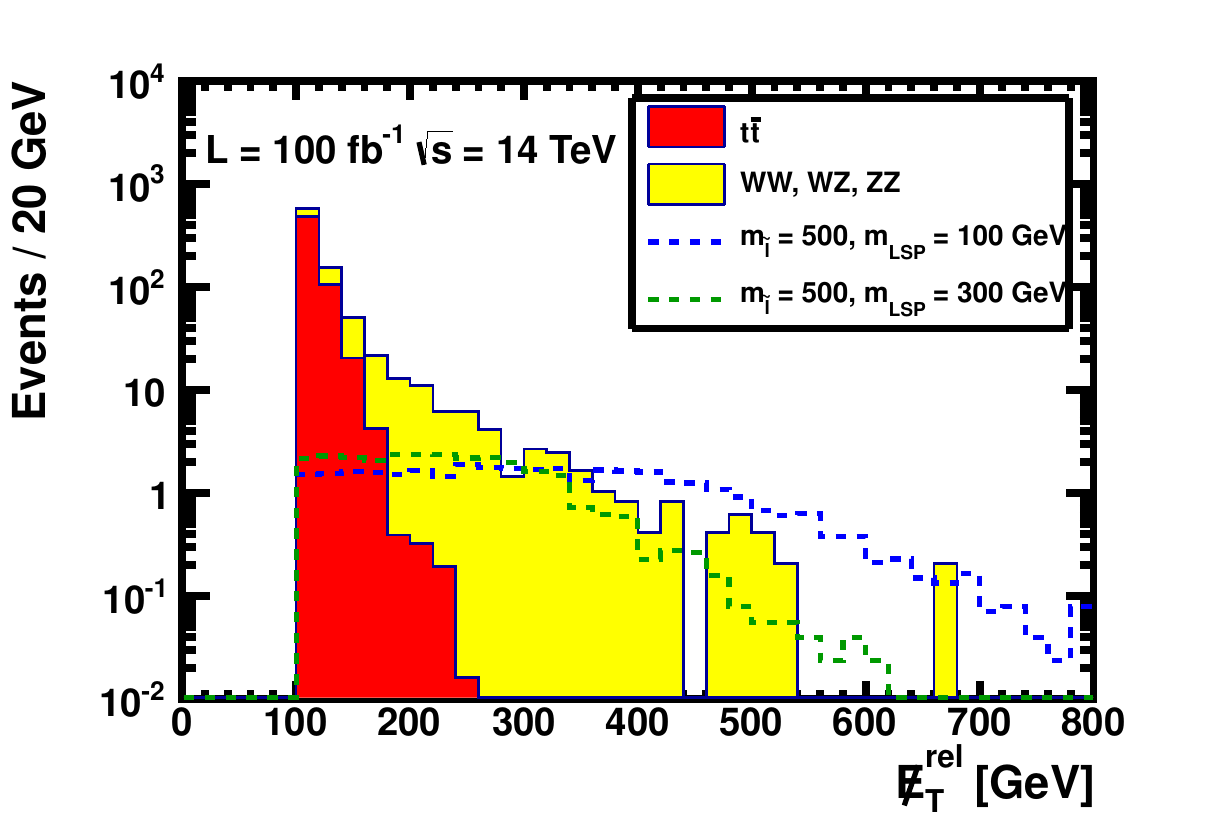}
 \minigraph{7.8cm}{-0.2in}{(b)}{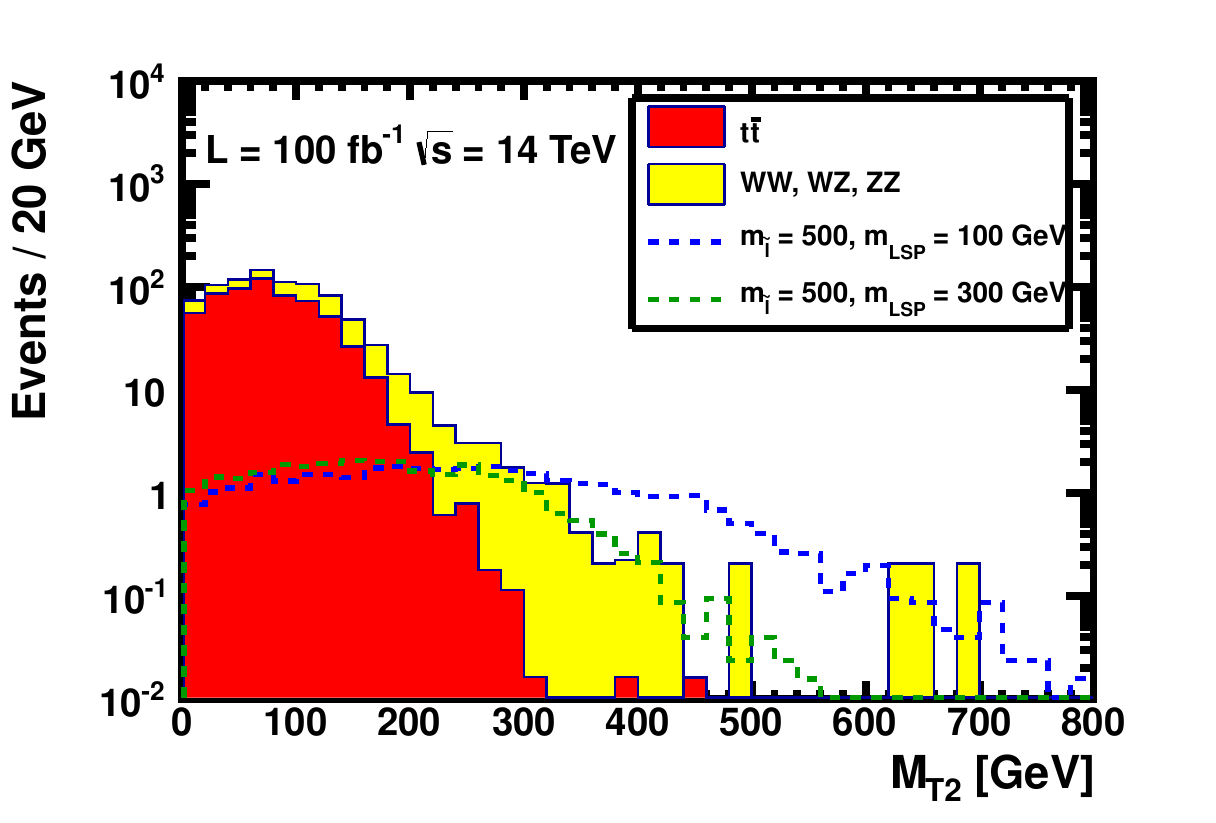}
\end{center}
\caption{Signal and background distributions for  (a) $\met^{{\rm rel}}$  and (b) $M_{T2}$  for 100 ${\rm fb}^{-1}$ integrated luminosity at the 14 TeV LHC.   Selection cuts, jet veto,  $Z$ veto, and $\met^{{\rm rel}}$ cut of 100 GeV have been imposed.
The signal distributions are shown for two benchmark points   of ($m_{\tilde\ell_L}$, $m_{\tilde\chi_1^0}$) = (500, 100) GeV and ($m_{\tilde\ell_L}$, $m_{\tilde\chi_1^0}$) = (500, 300) GeV with a Bino-like LSP.}
\label{fig:14TeV_metrel_MT2}
\end{figure}

\begin{table}
\begin{center}
\begin{tabular}{|l|c|c|c|c|c|c|}
\hline 
&Signal&&$t\bar{t}$&&Di-boson& \\
&$ee$&$\mu\mu$&$ee$&$\mu\mu$&$ee$&$\mu\mu$ \\
\hline
CS [fb]&$3.2 \times 10^{-1}$&$3.2 \times 10^{-1}$&$4.0 \times 10^{3}$&$4.0 \times 10^{3}$&$8.3 \times 10^{2}$&$8.3 \times 10^{2}$\\ \hline
 Selection Cuts &80\%&82\%&14\%&16\%&12\%&14\%\\
Jet Veto &64\%&66\%&4\%&4\%&10\%&11\%\\
$Z$ Veto &64\%&65\%&4\%&4\%&9\%&10\%\\
$M_{T2} > 50~\gev$&51\%&52\%&1\%&1\%&2\%&2\%\\
 $\met^{{\rm rel}} > 150~\gev$&33\%&34\%&$<$ 0.01\%&$<$ 0.01\%&0.03\%&0.02\%\\ \hline
CS after cuts [fb] &$1.1 \times 10^{-1}$&$1.1 \times 10^{-1}$&$1.0 \times 10^{-5}$&$1.7 \times 10^{-5}$&$2.2 \times 10^{-1}$&$1.6 \times 10^{-1}$\\ \hline
 $S/\sqrt{B}$ $@~100~\rm{fb}^{-1}$&  & combined: & 3.5 & & &\\
\hline
\end{tabular}
\end{center}
\caption{Cut efficiency for benchmark point of ($m_{\tilde\ell_L}$, $m_{\tilde\chi_1^0}$) = (500, 100) GeV with a Bino-like LSP, using cuts specified above.  Signal significance is shown for the combined $ee$ and $\mu \mu$ channels using 100 $\rm{fb}^{-1}$ of data at the 14 TeV LHC.  The $t\bar{t}$ cross section before the selection cuts already include a precut of  $p_T^j<100$ GeV.  }
\label{tab:eff} 
\end{table}

In Fig.~\ref{fig:14TeV_metrel_MT2}, we show the $\met^{{\rm rel}}$ and $M_{T2}$ distributions for both the backgrounds and two signal benchmark points  of ($m_{\tilde\ell_L}$, $m_{\tilde\chi_1^0}$) = (500, 100) GeV and ($m_{\tilde\ell_L}$, $m_{\tilde\chi_1^0}$) = (500, 300) GeV with a Bino-like LSP  after imposing the selection cuts, jet and $Z$ vetoes, and minimum $\met^{{\rm rel}}$ cut of 100 GeV. We observe that the backgrounds and signal (for the larger slepton-LSP mass splitting) become comparable at $\met^{{\rm rel}}$ and $M_{T2}$ on the order of 300 GeV, na\"ively suggesting imposing cuts in the vicinity of this region. However, the background distributions tend to fall quickly at higher $\met^{{\rm rel}}$ and $M_{T2}$, so to obtain sufficient statistics we impose somewhat looser cuts, with the requirements on $B$, $S$ and $S/B$ listed above.   The resulting, illustrative cut efficiencies are listed in Table \ref{tab:eff} for the benchmark point of ($m_{\tilde\ell_L}$, $m_{\tilde\chi_1^0}$) = (500, 100) GeV with a Bino-like LSP.

\begin{figure}[h]
  \minigraph{7.8cm}{-0.2in}{(a)}{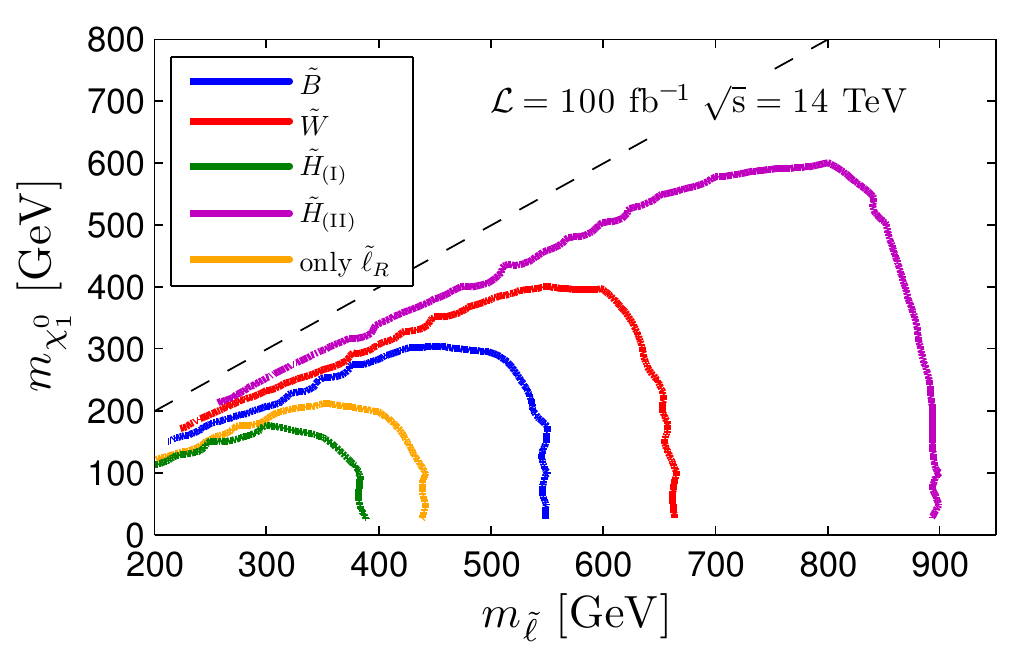}
\minigraph{7.8cm}{-0.2in}{(b)}{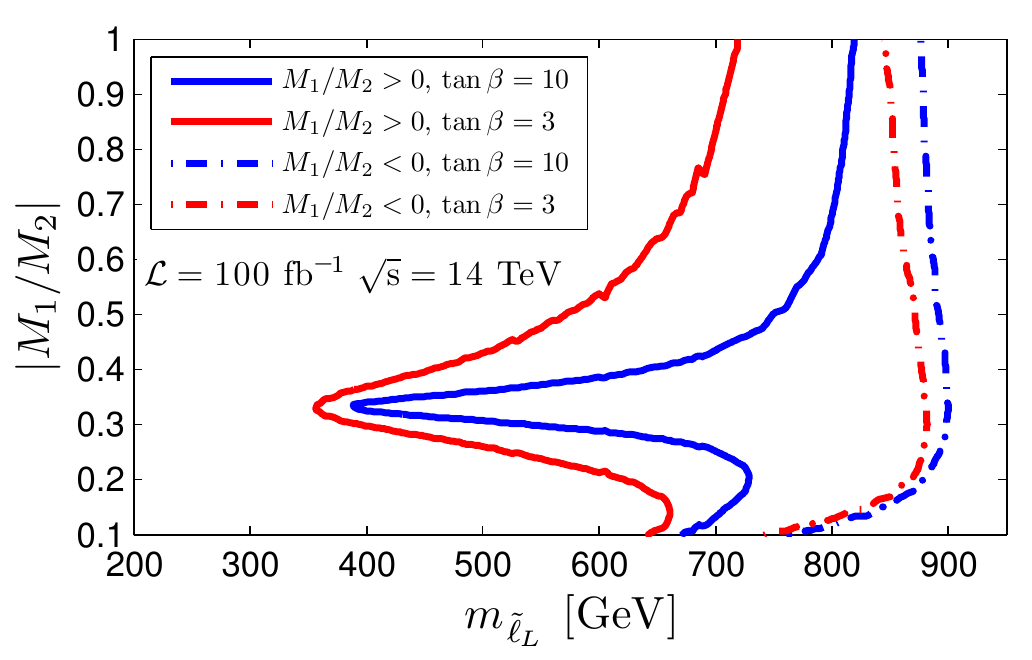}
  \caption{ Prospective 95\% C.L. exclusion limits  in the (a) $m_{\tilde\ell} - m_{\chi_1^0}$ plane, (b) $m_{\tilde\ell_L} - |M_1/M_2|$ plane  for slepton pair  production with dilepton $+\met$ final states, with 100  ${\rm fb}^{-1}$  luminosity at the 14 TeV LHC for various LSP scenarios.  The color coding and parameter choices  are the same as in Fig.~\ref{fig:8TeV_recast}.   }
\label{fig:14TeV_combined}
\end{figure}

The resulting, prospective 95\% C.L. expected exclusion limits  in the $m_{\tilde\ell} - m_{\chi_1^0}$ plane are given in Fig.~\ref{fig:14TeV_combined} (a)  for the 14 TeV LHC with 100  ${\rm fb}^{-1}$  integrated luminosity  for various slepton and neutralino LSP scenarios.    For the right-handed slepton, the reach is about 430 GeV for small LSP masses.  For the left-handed sleptons, the reach is 550 GeV (670 GeV) for the Bino-like (Wino-like) LSP case, and about 400 GeV 
and 900 GeV for the Higgsino-like LSP cases (I) and (II), respectively.    We find the reach with 300 ${\rm fb}^{-1}$ is typically about  50 $-$ 100 GeV better.  A 5\% systematic error  has been included in our limits to give a reasonably realistic reach for the LHC. 
 
The prospective 14 TeV exclusion reach in the $m_{\tilde\ell} - |M_1/M_2|$ plane for the Higgsino-like LSP case is shown in Fig.~\ref{fig:14TeV_combined} (b) for 100 ${\rm fb}^{-1}$  integrated luminosity. Regions to the left-side of the curves are excluded.  The weakest reach is for the $M_1/M_2>0$ case with small $\tan\beta$.  In particular, for $M_1/M_2 \sim \tan^2\theta_W$, the slepton mass reach is only about 350 GeV for $\tan\beta=3$ with 100 ${\rm fb}^{-1}$ luminosity.  The slepton mass reach increases when $M_1/M_2$ deviates from $\tan^2\theta_W$, approaching about 650 GeV for small $M_1/M_2$ and 710 GeV for large $M_1/M_2$.   The slepton mass reach for negative $M_1/M_2$ is typically better, around  800 $-$ 900 GeV, a pattern similar to that found in the 8 TeV analysis.    Comparing with  Fig.~\ref{fig:8TeV_recast} (b) we observe that the presently allowed region for small $m_{\tilde\ell}$ in the Higgsino-like LSP scenario with  $M_1/M_2>0$ could be excluded with the higher energy run.  For $M_1/M_2<0$, the exclusion reach becomes as much as a factor of two stronger than at present with 300  ${\rm fb}^{-1}$ integrated luminosity.

\begin{figure}[h]
\minigraph{7.8cm}{-0.2in}{(a)}{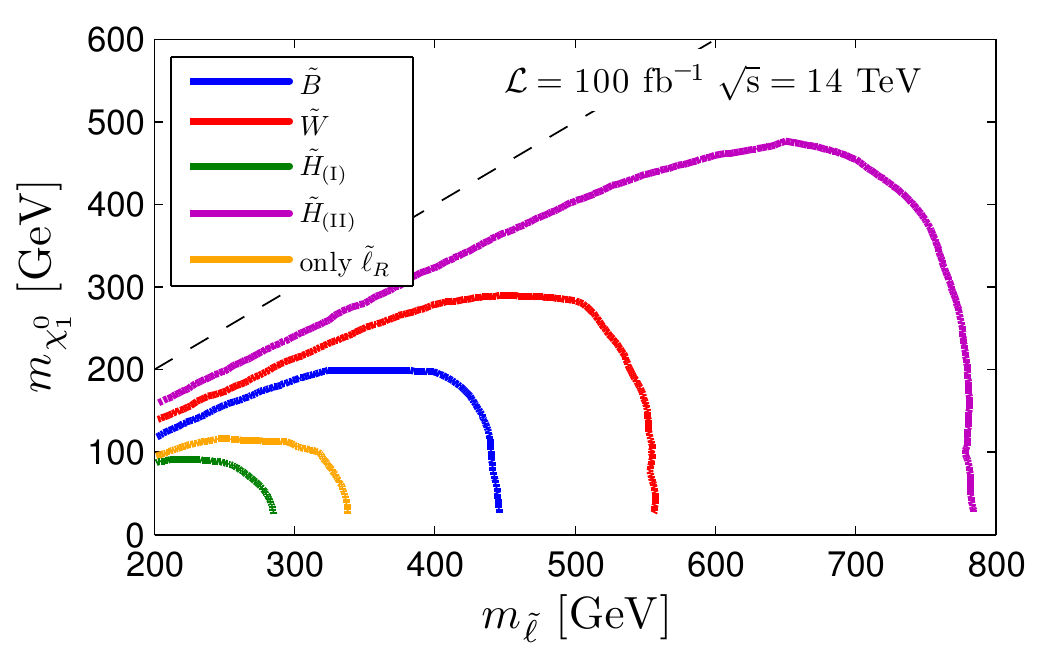}
\minigraph{7.8cm}{-0.2in}{(b)}{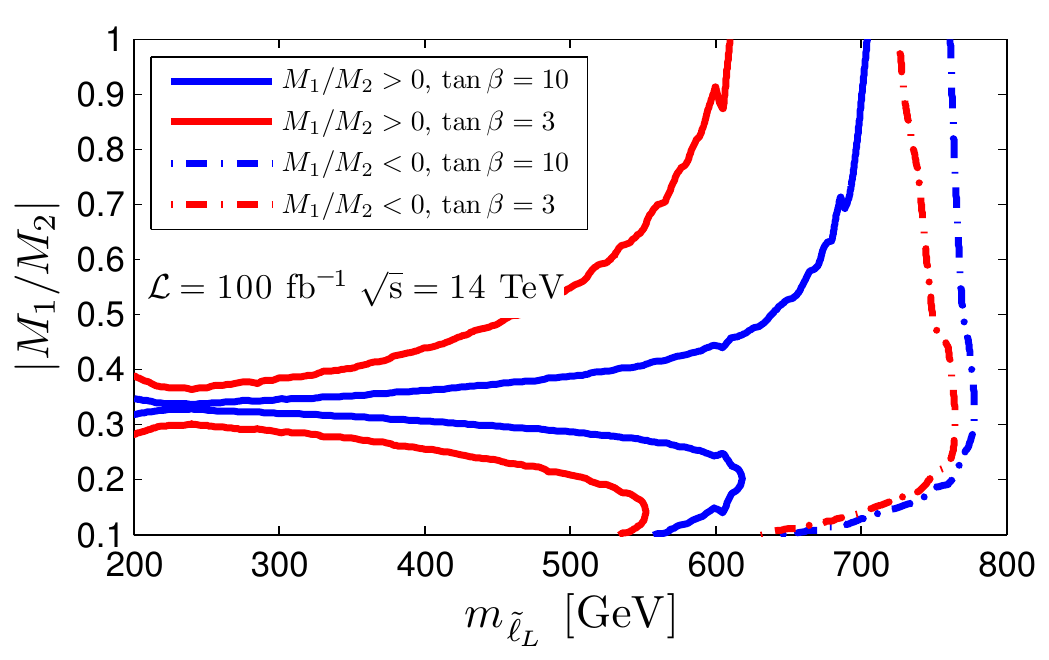}
  \caption{5$\sigma$ discovery   reach at the 14 TeV LHC  in the (a) $m_{\tilde\ell} - m_{\chi_1^0}$ plane,  (b) $m_{\tilde\ell_L} - |M_1/M_2|$ plane  for slepton pair  production with dilepton $+\met$ final states, with 100  ${\rm fb}^{-1}$  integrated  luminosity for various LSP scenarios.  The color coding and parameter choices are  the same as in Fig.~\ref{fig:8TeV_recast}.  
}
\label{fig:14TeV_combined_reach}
\end{figure}

 In Fig.~\ref{fig:14TeV_combined_reach} (a)  we show   the $5\sigma$ discovery reach for the various LSP benchmark scenarios we have considered previously.  The maximum reach occurs for the Higgsino-like LSP with $M_1/M_2=-1/3$, for which sleptons as heavy as  $\sim 800$ GeV  could be discovered with 100  ${\rm fb}^{-1}$  integrated luminosity. For a very light LSP, the reach is roughly three times weaker for $M_1/M_2=1/3$ case, while the reach for the Wino- and Bino-like LSP scenarios fall in between. Fig.~\ref{fig:14TeV_combined_reach} (b) gives the corresponding discovery potential in the $m_{\tilde\ell_L} - |M_1/M_2|$ plane for a Higgsino-like LSP.   While no sensitivity for slepton could be achieved for the worse case scenario of $M_1/M_2 \sim \tan^2\theta_W$, reaches in $m_{\tilde\ell}$ increases when $M_1/M_2$ deviates from this value.  For $M_1/M_2<0$, 5$\sigma$ reach can be as large as 800 GeV.   With 300 ${\rm fb}^{-1}$ at the 14 TeV LHC, the reach is improved by about  50 $-$ 100 GeV.

\section{Summary and Conclusion}
\label{sec:conclusion}

With the absence thus far of any superpartner signals at the LHC, the attention for LHC Run-II (14 TeV) will clearly require emphasis on more difficult-to-observe signatures. Among the most challenging are those associated with sleptons, given the $\mathcal{O}$(fb) electroweak production cross sections. In this work, we studied the dependence of slepton decay branching fractions on the nature of the LSP.  In particular, in the Higgsino-like LSP scenarios, both  decay branching fractions of $\tilde\ell_L$ and $\tilde\nu_\ell$ exhibit strong dependence on the sign and value of $M_1/M_2$: $\tilde\ell_L \rightarrow \ell \chi_{1,2}^0$  is minimized  for $M_1/M_2\sim   \tan^2\theta_W$, while  $\tilde\nu_\ell \rightarrow \ell \chi_{1}^\pm$  is maximized  for $M_1/M_2\sim -\tan^2\theta_W$.    Combined with the slepton pair production, we   analyzed the prospective reach for the OS dilepton plus $\met$ final state at the 8 and 14 TeV LHC.  

We recasted the existing 8 TeV results,  reported by the LHC collaborations assuming a Bino-like LSP,  in various LSP scenarios.   We find that the LHC slepton reach is strongly enhanced for a non-Bino-like LSP: the 95\% C.L. limit for $m_{\ell_L}$ extends from 300 GeV for Bino-like LSP to about 370 GeV for  Wino-like LSP.  More interestingly, the reach in the Higgsino-like LSP scenario sensitively depends on the value and sign of $M_1/M_2$.  The 95\% C.L. reach  for $\tilde\ell_L$ is the strongest ($\sim$ 490 GeV) for $M_1/M_2\sim -\tan^2\theta_W$ and is the weakest ($\sim$  220 GeV) for $M_1/M_2\sim \tan^2\theta_W$.

We also studied  the 95\% C.L. exclusion and 5$\sigma$ discovery reach of slepton at the 14 TeV LHC with 100 fb$^{-1}$ luminosity.  The projected 95\% C.L. mass limits for the left-handed slepton varies from 550 (670) GeV for a Bino-like (Wino-like) LSP to 900 (390) GeV for a Higgsino-like LSP under the most optimistic (pessimistic) scenario. The reach for the right-handed slepton is about 440 GeV.  The corresponding 5$\sigma$ discovery is about 100 GeV smaller.     For 300 $\rm fb^{-1}$ integrated luminosity, the reach is about 50 $-$ 100 GeV higher.

Interestingly, relatively light leptons with moderate $\tan\beta$ are needed to explain the present difference between the muon anomalous magnetic moment experimental result and the SM prediction.  The LHC Run-II should, thus, be able to probe this possibility for the Wino-like and Higgsino-like LSP.  The observation of a signal in this case could be consistent with a supersymmetric explanation for the $g_\mu-2$ result\footnote{The various one-loop contributions to the anomalous magnetic moment are proportional to $\mathrm{sign}(\mu M_j)$ for $j=1,2$, depending on which neutralino or chargino appears in the loop\cite{muong-2}. Thus, knowing the relative sign of $M_1$ and $M_2$, as well as the values of the superpartner masses, will allow for a precise determination of the MSSM contribution to the anomalous magnetic moment.}. In addition, one may also expect signatures in other low-energy electroweak processes, such as tests of lepton universality with pion leptonic decays or deviations from first row CKM unitarity as probed by $\beta$-decay and kaon leptonic decays. On the other hand, the non-observation of dilepton plus $\met$ signal for slepton Drell-Yan pair production   would not generally preclude light sleptons, as the rates for the right-handed  sleptons and for the left-handed  sleptons with a Higgsino-like LSP and $M_1/M_2 \sim \tan^2\theta_W$ are considerably suppressed. Probing these MSSM scenarios would require alternate avenues, such as the production of sleptons via the cascade decays from electroweak gaugino production or future studies at a high energy $e^+e^-$ collider.

\acknowledgments
 The work was supported in part under U.S. Department of Energy contracts DE-FG02-04ER-41298 (S.S. and J.E.), DE-FG02-04ER-41268 (W.S.), 
DE-FG02-08ER41531 (M.J.R-M) and DE-SC0011095 (M.J.R-M) as well as by the Wisconsin Alumni Research Foundation (M.J.R-M).


\begin{thebibliography}{99}



 
 
 \bibitem{aad:2012gk} 
  G.~Aad {\it et al.}  [ATLAS Collaboration],
  Phys.\ Lett.\ B {\bf 716}, 1 (2012)
  [arXiv:1207.7214 [hep-ex]].

  


\bibitem{Chatrchyan:2012ufa} 
  S.~Chatrchyan {\it et al.}  [CMS Collaboration],
  Phys.\ Lett.\ B {\bf 716}, 30 (2012)
  [arXiv:1207.7235 [hep-ex]].

\bibitem{SUSYcolor}
  G.~Aad {\it et al.}  [ATLAS Collaboration],
  ATLAS-CONF-2013-047;
  S.~Chatrchyan {\it et al.}   [CMS Collaboration],
  CMS-PAS-SUS-13-012.

\bibitem{GMSB} For a review, see 
  G.~F.~Giudice, R.~Rattazzi,
  Phys.\ Rept.\  {\bf 322}, 419-499 (1999) 
  [hep-ph/9801271].

\bibitem{AMSB} 
  L.~Randall, R.~Sundrum,
  Nucl.\ Phys.\  {\bf B557}, 79-118 (1999) 
  [hep-th/9810155];
  G.~F.~Giudice, M.~A.~Luty, H.~Murayama, R.~Rattazzi,
  JHEP {\bf 9812}, 027 (1998)
  [hep-ph/9810442];
  T.~Gherghetta, G.~F.~Giudice, J.~D.~Wells,
  Nucl.\ Phys.\  {\bf B559}, 27-47 (1999)
  [hep-ph/9904378].
  
  
  

\bibitem{mSUGRA} 
For a review, see 
  A.~B.~Lahanas, D.~V.~Nanopoulos,
  Phys.\ Rept.\  {\bf 145}, 1 (1987).

\bibitem{neutralinoDM} 
  H.~Goldberg,
  Phys.\ Rev.\ Lett.\  {\bf 50}, 1419 (1983);
  J.~R.~Ellis, J.~S.~Hagelin, D.~V.~Nanopoulos, K.~A.~Olive, M.~Srednicki,
  Nucl.\ Phys.\  {\bf B238}, 453-476 (1984).
  


\bibitem{sleptonrelic}M.~Drees, M.~M.~Nojiri,
  Phys.\ Rev.\  {\bf D47}, 376-408 (1993)
  [hep-ph/9207234];
  T.~Nihei, L.~Roszkowski, R.~Ruiz de Austri,
  JHEP {\bf 0203}, 031 (2002)
  [hep-ph/0202009];
  A.~Birkedal-Hansen, E.~-h.~Jeong,
  JHEP {\bf 0302}, 047 (2003)
  [hep-ph/0210041].

\bibitem{coann} 
  K.~Griest and D.~Seckel,
  Phys.\ Rev.\ D {\bf 43}, 3191 (1991).

\bibitem{moller} 
  J.~Mammei [MOLLER Collaboration],
  Nuovo Cim.\ C {\bf 035N04}, 203 (2012)
  [arXiv:1208.1260 [hep-ex]].

\bibitem{qweak} 
  D.~Androic {\it et al.}  [Qweak Collaboration],
  Phys.\ Rev.\ Lett.\  {\bf 111}, no. 14, 141803 (2013)
  [arXiv:1307.5275 [nucl-ex]].

\bibitem{muong-2}
  T.~Moroi,
  Phys.\ Rev.\ D {\bf 53}, 6565 (1996)
  [Erratum-ibid.\ D {\bf 56}, 4424 (1997)]
  [hep-ph/9512396].

\bibitem{CKM}
  V.~Cirigliano, J.~Jenkins and M.~Gonzalez-Alonso,
  Nucl.\ Phys.\ B {\bf 830}, 95 (2010)
  [arXiv:0908.1754 [hep-ph]].

\bibitem{Kumar:2013qya} 
  K.~Kumar, Z.~-T.~Lu and M.~J.~Ramsey-Musolf,
  arXiv:1312.5416 [hep-ph].
  
\bibitem{Albrecht:2013wet} 
  J.~Albrecht {\it et al.}  [Intensity Frontier Charged Lepton Working Group Collaboration],
  arXiv:1311.5278 [hep-ex].




\bibitem{delAguila:1990yw} 
  F.~del Aguila and L.~Ametller,
  Phys.\ Lett.\ B {\bf 261}, 326 (1991).
  
\bibitem{Baer:1993ew} 
  H.~Baer, C.~-h.~Chen, F.~Paige and X.~Tata,
  Phys.\ Rev.\ D {\bf 49}, 3283 (1994)
  [hep-ph/9311248].
  
\bibitem{Andreev:2004qq} 
  Y.~.M.~Andreev, S.~I.~Bityukov and N.~V.~Krasnikov,
  Phys.\ Atom.\ Nucl.\  {\bf 68}, 340 (2005)
  [Yad.\ Fiz.\  {\bf 68}, 366 (2005)]
  [hep-ph/0402229].

\bibitem{sleptonATLAS}
  G.~Aad {\it et al.}  [ATLAS Collaboration],
  ATLAS-CONF-2013-049.

\bibitem{sleptonCMS}
  S.~Chatrchyan {\it et al.}   [CMS Collaboration],
  CMS-PAS-SUS-13-006.

\bibitem{Aad:2014vma} 
  G.~Aad {\it et al.}  [ATLAS Collaboration],
  JHEP {\bf 1405}, 071 (2014)
  [arXiv:1403.5294 [hep-ex]].
  
\bibitem{Khachatryan:2014qwa} 
  V.~Khachatryan {\it et al.}  [CMS Collaboration],
  arXiv:1405.7570 [hep-ex].

\bibitem{Eckel:2011pw} 
  J.~Eckel, W.~Shepherd and S.~-F.~Su,
  JHEP {\bf 1205}, 081 (2012)
  [arXiv:1111.2615 [hep-ph]].



\bibitem{sleptonflavor} 
J.~L.~Feng, S.~T.~French, I.~Galon, C.~G.~Lester, Y.~Nir, Y.~Shadmi, D.~Sanford, F.~Yu,
  JHEP {\bf 1001}, 047 (2010)
  [arXiv:0910.1618 [hep-ph]];
  J.~L.~Feng, S.~T.~French, C.~G.~Lester, Y.~Nir, Y.~Shadmi,
  Phys.\ Rev.\  {\bf D80}, 114004 (2009)
  [arXiv:0906.4215 [hep-ph]].
  
\bibitem{Krasnikov:1996np}
  N.~V.~Krasnikov,
  JETP Lett.\  {\bf 65}, 148-153 (1997)
  [hep-ph/9611282].





 




 \bibitem{LEPslepton}
 LEP2 SUSY Working Group,  ``Combined LEP Selectron/Smuon/Stau Results, 183-208 GeV", (2004),
 note LEPSUSYWG/04-01.1, 
 http://lepsusy.web.cern.ch/lepsusy/.
 
\bibitem{LEPsinglelep}
  P.~Achard {\it et al.} [ L3 Collaboration ],
  Phys.\ Lett.\  {\bf B580}, 37-49 (2004)
  [hep-ex/0310007].
  A.~Heister {\it et al.} [ ALEPH Collaboration ],
  Phys.\ Lett.\  {\bf B544}, 73-88 (2002)
  [hep-ex/0207056].  
  
\bibitem{:2005ema}
  [ ALEPH and DELPHI and L3 and OPAL and SLD and LEP Electroweak Working Group and SLD Electroweak Group and SLD Heavy Flavour Group Collaborations ],
  Phys.\ Rept.\  {\bf 427}, 257-454 (2006)
  [hep-ex/0509008].


\bibitem{Alwall:2011uj} 
  J.~Alwall, M.~Herquet, F.~Maltoni, O.~Mattelaer and T.~Stelzer,
  JHEP {\bf 1106}, 128 (2011)
  [arXiv:1106.0522 [hep-ph]].

\bibitem{Sjostrand:2006za} 
  T.~Sjostrand, S.~Mrenna and P.~Z.~Skands,
  JHEP {\bf 0605}, 026 (2006)
  [hep-ph/0603175].
 
\bibitem{deFavereau:2013fsa} 
  J.~de Favereau, C.~Delaere, P.~Demin, A.~Giammanco, V.~Lemaître, A.~Mertens and M.~Selvaggi,
  arXiv:1307.6346 [hep-ex].

\bibitem{snowmass}
 http://www.snowmass2013.org/tiki-index.php?page=Energy\_Frontier\_FastSimulation


\bibitem{Fuks:2013lya} 
  B.~Fuks, M.~Klasen, D.~R.~Lamprea and M.~Rothering,
  JHEP {\bf 1401}, 168 (2014)
  [arXiv:1310.2621, arXiv:1310.2621 [hep-ph]].

\bibitem{Lester:1999tx} 
  C.~G.~Lester and D.~J.~Summers,
  Phys.\ Lett.\ B {\bf 463}, 99 (1999)
  [hep-ph/9906349].

\bibitem{Barr:2003rg} 
  A.~Barr, C.~Lester and P.~Stephens,
  J.\ Phys.\ G {\bf 29}, 2343 (2003)
  [hep-ph/0304226].

\bibitem{Cheng:2008hk} 
  H.~-C.~Cheng and Z.~Han,
  JHEP {\bf 0812}, 063 (2008)
  [arXiv:0810.5178 [hep-ph]].

\bibitem{Junk:1999kv} 
  T.~Junk,
  Nucl.\ Instrum.\ Meth.\ A {\bf 434}, 435 (1999)
  [hep-ex/9902006].

\bibitem{Read:2000ru} 
  A.~L.~Read,
  In {\it Geneva 2000, Confidence limits},  81-101.

\bibitem{Bern:2008ef} 
  Z.~Bern {\it et al.}  [NLO Multileg Working Group Collaboration],
  arXiv:0803.0494 [hep-ph].




 
\end{thebibliography}
\end{document}